\def\year{2020}\relax
\documentclass[letterpaper]{article} 
\usepackage{aaai20}  
\usepackage{times}  
\usepackage{helvet} 
\usepackage{courier}  
\usepackage[hyphens]{url}  
\usepackage{graphicx} 
\usepackage{graphicx}  
\usepackage{url}  
\usepackage{graphicx}  

\usepackage{amsmath,amssymb}
\usepackage{subfig}
\usepackage{xcolor}

\usepackage{tikz}
\usetikzlibrary{arrows}
\usetikzlibrary{bayesnet}
\usepackage{multirow}

\usepackage{algorithm}
\usepackage{algpseudocode}
\usepackage{bbm}
\usepackage{bm}

\usepackage{enumitem}
\usepackage{color, colortbl}
\usepackage{lipsum}
\definecolor{Gray}{gray}{0.935}
\usepackage{booktabs}

\usepackage{makecell}

\newcommand{\xhdr}[1]{\vspace{1.0mm}\noindent{{\bf #1.}}}
\newcommand{\xhqr}[1]{\vspace{1.0mm}\noindent{{\bf #1?}}}

\title{Engagement Patterns of Peer-to-Peer Interactions on Mental Health Platforms}
\newcommand\aspace{\hspace{0.4em}}

\newcommand{\citet}[1] {\citeauthor{#1}~\shortcite{#1}}

\author{Ashish Sharma\textsuperscript{\rm 1}\thanks{This work was done when the author was a Research Fellow at Microsoft Research, India.} \aspace{} Monojit Choudhury\textsuperscript{\rm 2} \aspace{} Tim Althoff\textsuperscript{\rm 1} \aspace{} Amit Sharma\textsuperscript{\rm 2} \vspace*{.3em} \\
\textsuperscript{\rm 1}Paul G. Allen School of Computer Science \& Engineering, University of Washington, Seattle, USA \vspace*{.15em} \\
\textsuperscript{\rm 2}Microsoft Research, Bangalore, India \vspace*{.3em}  \\
\textsuperscript{\rm 1}\texttt{\{ashshar,althoff\}@cs.washington.edu} \aspace{} \textsuperscript{\rm 2}\texttt{\{monojitc,amshar\}@microsoft.com}}

\begin{document}

\maketitle

\begin{abstract}
Mental illness is a global health problem, but access to mental healthcare resources remain poor worldwide. Online peer-to-peer support platforms attempt to alleviate this fundamental gap by enabling those who struggle with mental illness to provide and receive social support from their peers. However, successful social support requires users to engage with each other and failures may have serious consequences for users in need. Our understanding of engagement patterns on mental health platforms is limited but critical to inform the role, limitations, and design of these platforms. Here, we present a large-scale analysis of engagement patterns of 35 million posts on two popular online mental health platforms, \textsc{TalkLife} and \textsc{Reddit}. Leveraging communication models in human-computer interaction and communication theory, we operationalize a set of four engagement indicators based on attention and interaction. We then propose a generative model to jointly model these indicators of engagement, the output of which is synthesized into a novel set of eleven distinct, interpretable patterns. We demonstrate that this framework of engagement patterns enables informative evaluations and analysis of online support platforms. Specifically, we find that mutual back-and-forth interactions are associated with significantly higher user retention rates on~\textsc{TalkLife}. Such back-and-forth interactions, in turn, are associated with early response times and the sentiment of posts.
\end{abstract}

\section{Introduction}

Mental illness is an alarming global health issue with adverse social and economic consequences. Mental illness and related behavioral health problems contribute 13\% to the global burden of disease, more than cardiovascular diseases and cancer~\cite{collins2011grand}. Still, access to mental health care is poor worldwide. Most low-income and middle-income countries have less than one psychiatrist per 100,000 individuals~\cite{rathod2017mental}. Even in high-income countries like the United States, 60\% of counties do not have a single psychiatrist (New-American-Economy Research, 2019). 

Research suggests that for people in distress, connecting and interacting with \textit{peers} can be helpful in coping with mental illness, enhancing mental well-being and developing social integration~\cite{davidson1999peer}. This form of social support~\cite{kaplan1977social} through peers can be provided online which has stimulated the design and development of online mental health support platforms. 

In recent years, several low-cost and easy-to-access peer-to-peer support platforms, such as \textsc{TalkLife} \& 7Cups\footnote{\url{https://talklife.co/}, \url{https://www.7cups.com/}}, have provided new pathways for seeking social support and dealing with mental health challenges. These platforms allow interactions between support seekers and peers in a thread-like setting; it starts with a user writing a support seeking post which elicits responses from peers and subsequent interactions between the users. Online platforms have multiple advantages over traditional face-to-face supportive methods: they enable asynchronous conversations by design; they are unrestricted by time, space and geographic boundaries; and they facilitate anonymous disclosures which can be helpful in dealing with the major challenge of stigma associated with mental illness~\cite{white2001receiving}.

However, for these platforms to be successful at facilitating peer-to-peer support, users need to interact and engage. A user who wants to seek support on the platform (henceforth referred to as \textit{seeker}) needs to interact with a peer who is willing to provide support (henceforth referred as \textit{peer-supporter}). 
For example, on \textsc{TalkLife}, one third of support-seeking posts by users do not receive any responses at all. 
Receiving no response or having limited engagement with peers can have serious consequences on a mental health platform with vulnerable users, a number of whom are at risk for self-harm or suicide. Also, as indicated in prior literature, engagement between users is key for ensuring favorable outcomes on these platforms~\cite{van2011determinants}, including overcoming cognitive distortion, effective distraction, and empathy~\cite{taylor2011avatars,mayshak2017influence}.

Prior work on engagement between users on support platforms have focused on finding its correlations with several user and platform related characteristics, such as methods of support seeking~\cite{andalibi2018social}, support providing~\cite{andalibi2018responding}, and self-disclosure~\cite{ernala2018characterizing}. However, these works have either been conducted as user studies or studies over small human-annotated datasets, or have made strong assumptions {in their characterizations of engagement by overlooking factors  such as the degree of interaction between users, that are key to engagement in conversations as noted in communication theory~\cite{bretz1983media,williams1988research,sheizaf1988interactivity}.} Furthermore, none of them attempt to {develop a collective sense of engagement}; instead, they independently {examine} various engagement dimensions, such as the number of posts and likes. 

\xhdr{Present Work} In this paper, we conduct a large-scale study of thread-level engagement patterns of 35 million posts across 8 million threads on~\textsc{TalkLife} and \textsc{Reddit}. We take a microscopic view of engagement between users on the platform and focus on engagement at the level of conversational thread. Drawing inspirations from Human-Computer-Interaction~\cite{o2008user} and Communication theory~\cite{bretz1983media,williams1988research,rosengren1999communication,ridley1979social}, we operationalize a set of quantitative \textit{indicators} of thread-level engagement  around the notions of attention {(the amount of attention received by a thread)}, and interaction {(the nature of interaction between the users in the thread)} (Section~\ref{sec:indicators}). 
We demonstrate that no single engagement indicator can fully capture observated engagement dynamics. Therefore, we jointly model multiple engagement indicators and discover interpretable thread-level \textit{engagement patterns}.
We design a generative model which learns distinct clusters of engagement patterns as a probability distribution over the joint space of multiple engagement indicators (Section~\ref{sec:model}). We analyze these clusters to derive a set of 11 novel, interpretable engagement patterns (Section~\ref{sec:taxonomy}). 

We demonstrate that our novel framework of engagement patterns enables online support platforms to conduct informative self-evaluations and comparative assessments (Section~\ref{sec:insights}). For example, an analysis of \textsc{TalkLife} using the framework informs us that a mutual discourse (back-and-forth interactions; Figure~\ref{subfig:fully-interactive}) between seekers and peer-supporters is more important for seeker retention than all other engagement indicators. Such an insight is critical for the platforms; platform designers need to uncover design techniques that enable mutual interactions. Moreover, a comparative analysis using our framework highlights the impact of design differences between \textsc{TalkLife} \& \textsc{Reddit} to engagement dynamics between users. We end with a discussion of the limitations and risks of our findings for designing mental health support interventions (Section~\ref{sec:discussion}).

\section{Related Work}
Our work builds upon the studies of engagement in online communities, research on social support for mental illness and the design of statistical methods for modeling threads.

\subsection{Engagement patterns in online platforms}
The notion of engagement is a complex amalgamation of varied facets; there is no clear way of defining engagement~\cite{smith2017trajectories}. The definitions of engagement are adapted based on its context of usage with the focus being on attention, interaction, and affective experience~\cite{o2008user}. Thus, researchers commonly use various context-specific \textit{markers} or \textit{indicators} of engagement. In a recent work on a sexual abuse subreddit, Andalibi et al.~\shortcite{andalibi2018social} used the length of the thread as the sole indicator of engagement. They found that users who seek direct support receive more replies than the users who do not. 
Ernala et al.~\shortcite{ernala2018characterizing} studied the effect of responder's engagement on disclosures of highly stigmatized mental illnesses by users on Twitter. They used number of retweets, favorites \& mentions on Twitter as their engagement indicators and found positive correlations with the future intimacy of disclosures. Choudhury et al.~\shortcite{de2013predicting} studied the engagement of Twitter users before they are diagnosed with depression and used three attention-related engagement indicators -- number of posts, number of replies and retweets in response, and two content-related indicators {--} number of links shared by the user and number of question-centric posts. There has also been work on qualitatively analyzing engagement on MOOC forums. Mak et al.~\shortcite{mak2010blogs} in their qualitative framework and analysis of MOOC forums differentiate between threads with long-loops (slow single-user posts) and short-loops (quick multi-user posts).

In this work, we focus on thread-level engagement. Our work builds upon prior research by exploring two new interaction-based indicators of engagement (including a novel indicator of degree of interaction based on Communication theory~\cite{bretz1983media,williams1988research}) which, to the best of our knowledge, have not yet been used in the context of online social support and have limited research in other domains. Moreover, we jointly model and explore these indicators. The use of new indicators and their joint modeling provides us additional insights on the patterns of engagements and their associations with the user behavior on online support platforms. 
\subsection{Mental illness \& online social support} 
There is a rich body of work on detecting and diagnosing mental illness from posts and activities of users on social media platforms (Twitter~\cite{de2013predicting,coppersmith2014measuring,lin2014user,tsugawa2015recognizing}; Facebook~\cite{de2014characterizing}; Instagram~\cite{reece2017instagram}). A similar line of research is focused on estimating the severity of suicidal ideation and risk among individuals disclosing mental illness~\cite{de2016discovering,benton2017multitask,gaur2019knowledge}. Researchers have also made efforts in differentiating between the various types of support (e.g. \textit{informational}, \textit{emotional}) provided online ~\cite{biyani2014identifying} and analyzing {their effects} on suicidal risk~\cite{de2017language} and mental well-being~\cite{saha2020causal}. Studies analyzing self-disclosure~\cite{de2014mental,yang2019channel}, anonymity~\cite{de2014mental}, reciprocity~\cite{yang2019channel}, linguistic accommodation~\cite{sharma2018mental} and cognitive restructuring~\cite{pruksachatkun2019moments} on online support forums have also been conducted. {There has also been work} on analyzing the quality of online counseling conversations~\cite{perez2019makes} and studying their associations with conversation outcomes~\cite{althoff2016large}.

Our work is directed towards discovering patterns of effective online mental health support conversations. We focus on engagement in support conversations and develop a novel framework that enables informative evaluations and comparative assessments of mental health platforms.

\subsection{Modeling of a conversational thread}
A large set of online platforms such as Reddit, Twitter, etc. facilitate conversations in a thread-like setting. A user on the platform starts a thread which then elicits responses from other users along with back-and-forth interactions between users. This thread-like structure is typically modeled in a generative manner with the focus being on learning the growth dynamics of the thread (e.g. length) or the structural properties of the threads. Kumar et al.~\shortcite{kumar2010dynamics} modeled arrival of posts in a thread by incorporating both time and user of the post in a preferential attachment model. Wang et al.~\shortcite{wang2012user} used a continuous-time model over exposure duration and arrival rates of the posts to explain conversational growth. Backstrom et al.~\shortcite{backstrom2013characterizing} work on the tasks of length prediction and re-entry prediction of users in a thread using features related to post content, time, link and arrival patterns. They also make a distinction between \textit{focused} (long threads with a lot of comments from a small set of users) and \textit{expansionary} (long threads with few comments from large set of users) threads. Recently, Arag{\'o}n et al.~\shortcite{aragon2017thread} analyzed the differences between linear and hierarchical threads. Lumbreras et al. \shortcite{lumbreras2017role} propose a mixture model which learns latent roles of users; growth of the thread is dependent on the role of the users in the thread.

In this paper, we build on this literature 
by jointly modeling multiple engagement indicators and identifying clusters of engagement patterns. To this end, we design a generative model that discovers the desired meaningful clusters in an unsupervised setting (Section~\ref{sec:model}).


\section{Dataset Description}

\begin{center}
 \begin{table}[t]
 \centering
\begin{tabular}{p{0.3\columnwidth}p{0.27\columnwidth}p{0.27\columnwidth}}
\toprule

\textbf{Data Statistics} &  \textsc{TalkLife} & \textsc{Reddit} \\
\midrule
\rowcolor{Gray}
\# of Threads & 6.4M & 1.6M \\
\midrule
\# of Posts & 24.9M & 9.6M \\
\midrule
\rowcolor{Gray}
\# of Users & 339.4K & 969.7K \\
\midrule
Observation Period & May 2012 to Jan 2019 & Jan 2015 to Jan 2019 \\
\bottomrule
\end{tabular}
 \caption{Statistics of the two data sources.}
  \label{tab:dataset}
 \end{table}
\end{center}
We use conversational threads posted on two of the largest online support platforms as our data sources -- \textsc{TalkLife} (\url{talklife.co}) and mental health subreddits on \textsc{Reddit} (\url{reddit.com}). 

\xhdr{\textsc{TalkLife}} Founded in 2012, \textsc{TalkLife} is a free peer-to-peer network for mental health support. It enables people in distress to have interactions with other peers on the platform. One of the primary ways of having these interactions is using conversational threads which is the focus of our study. A \textit{conversational thread} or simply a \textit{thread} on \textsc{TalkLife} is characterized by a user initially authoring a post, typically seeking direct (e.g. \textit{I am struggling with thoughts of self-harm, someone please help}) or indirect (e.g. \textit{Life is like a miserable hassle!}) mental health support; the post then receives (un)supportive responses from the peers on the platform, sometimes leading to back-and-forth conversations between the users. We call the user who authors the first post to start the thread the \textit{seeker} and we call the users who post responses to the thread \textit{peer-supporters}. Note that the notions of seeker and peer-supporter are specific to a thread; a seeker may be a peer-supporter in a different thread. An alternative way of classifying users could be based on their time-aggregated platform activities~\cite{yang2019seekers}, which is beyond the scope of this work.

\xhdr{Mental Health Subreddits} \textsc{Reddit} is another popular online platform hosting conversational threads. It consists of a large number of sub-communities called \textit{subreddits}, each dedicated to a particular topic. We use threads posts on 55 mental health focused subreddits (list compiled by Sharma et al.~\shortcite{sharma2018mental}). We accessed the archive of reddit threads hosted on Google BigQuery spanning 2015 to 2019.  

\xhdr{\textsc{TalkLife} vs. \textsc{Reddit}} \textsc{Reddit} and \textsc{TalkLife} have a key difference in design. \textsc{Reddit} has topically-focused sub-communities which allows users to subscribe to topics they are most interested in. For example, users dealing with post traumatic stress disorder may only join \textit{r/ptsd}. As we show later, this difference may have major implications on conversational behavior and engagement dynamics on the two platforms (Section~\ref{subsec:compare}). In addition, only a small part of \textsc{Reddit} is focused on mental health-related interactions (less than 0.1\%) whereas all interactions on \textsc{TalkLife} are meant to be ideally focused on mental health. On both the platforms, however, mental health support is provided by volunteer peers (usually untrained) and rarely by professionals. 

Table~\ref{tab:dataset} summarizes the statistics of the two datasets. We make use of individual posts in the threads, their timestamps, and the IDs of the users who authored those posts. We minimize the use of additional metadata in our modeling that is highly specific to \textsc{TalkLife} or \textsc{Reddit} (e.g. tags of a thread) in order to promote methods \& results that could potentially generalize to other platforms.

\xhdr{Privacy, Ethics and Disclosure} The \textsc{TalkLife} dataset was sourced (with license and consent) from the \textsc{TalkLife} platform. All personally identifiable information was removed before analysis. In addition, all work was approved by Microsoft's Institutional Review Board. \textit{This work does not make any treatment recommendations or diagnostic claims}.

\xhdr{Data Access} The entire \textsc{Reddit} dataset used in this paper can be accessed from Google BigQuery\footnote{\url{https://bit.ly/2WQPosf}}. A sample of threads from the \textsc{TalkLife} dataset can be viewed online\footnote{\url{https://web.talklife.co/}} but should be used in accordance with their privacy policies and terms of service\footnote{\url{https://www.talklife.co/privacy}, \url{https://www.talklife.co/terms}}.


\section{Indicators of Thread-Level Engagement}
\label{sec:indicators}

Engagement cannot be fully captured by any single quantitative measure.
Instead, various context-specific \textit{markers} or \textit{indicators} are commonly used in engagement studies. In this paper, we are focused on taking a {microscopic} view of engagement on the mental health forums; we are interested in operationalizing \textit{thread-level engagement}. We look for indicators in a thread which would determine engagement between the seeker and the peer-supporters of the thread. 
We also keep our framework independent of the content of the individual posts; the content in mental health support platforms is often sensitive and may present ethical concerns. Finally, we want to have a collective and joint understanding of various dimensions of thread-level engagement (Section~\ref{subsec:model-motivation}) and thus, wish to use indicators which are complementary. Such a joint understanding is important for finding meaningful \textit{patterns of engagement} (Section~\ref{sec:taxonomy}).

\xhdr{Engagement vs. outcomes} While past work has focused on conversation outcomes (Althoff et al.~\shortcite{althoff2016large}, Pruksachatkun et al.~\shortcite{pruksachatkun2019moments}), here we focus on patterns of engagement. 
We note that a more engaging thread may not always be the more helpful thread. Depending on  context, there might be instances where a thread which is low in engagement is more helpful to a seeker than a highly engaging thread. 
Future work should investigate the link between engagement \& interaction outcomes.

We divide our engagement indicators into 2 categories -- \textit{attention-based} indicators and \textit{interaction-based} indicators. The attention-based indicators quantify the amount of attention received by the thread; interaction-based indicators quantify the nature of interaction between seekers \& peer-supporters in the thread. 

\subsection{Attention-Based Indicators}
\label{subsec:attention-indicator}
These indicators quantify the amount of attention a thread receives. We use the following indicators based on attention:

\xhdr{Thread Length} The number of posts (seeker posts and replies) in a thread. 
We aim to differentiate between threads with large (\textit{Long Threads}) and small (\textit{Short Threads}) number of posts. We observe that both \textsc{TalkLife} and \textsc{Reddit} contain a large number of threads of length = 1 (32.43\% \& 27.53\% respectively) along with a lot of short threads; long threads are low in proportion. We use generative modeling to determine appropriate thresholds of short and long threads. 




\xhdr{Peer-Supporters} The number of peer-supporters who post their replies to a thread. We contrast between threads having different engagement dynamics based on the number of peer-supporters, particularly between threads having individual and group communication dynamics~\cite{rosengren1999communication}. If a single peer-supporter responds on a thread, a direct communication takes place between the seeker and the peer-supporter (\textit{Two-Party Threads}). On the other hand, if multiple peer-supporters post on a thread, the communication happens in a group (\textit{Multi-Party Threads}). Moreover, we observe that a lot of threads on both \textsc{TalkLife} and \textsc{Reddit} receive no response at all; they have zero peer-supporters. We follow the terminology used by Ridley \& Avery~\shortcite{ridley1979social} and call them \textit{Isolated Threads}. 


\begin{figure}
\centering
\subfloat[Single Interaction]{
	\label{subfig:quasi-interactive}
	\includegraphics[width=0.32\columnwidth]{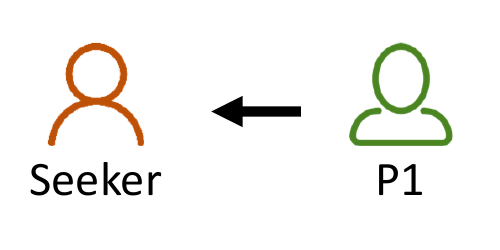} } 
\hfill
\subfloat[Repeated Seeker Interaction]{
	\label{subfig:interactive}
	\includegraphics[width=0.55\columnwidth]{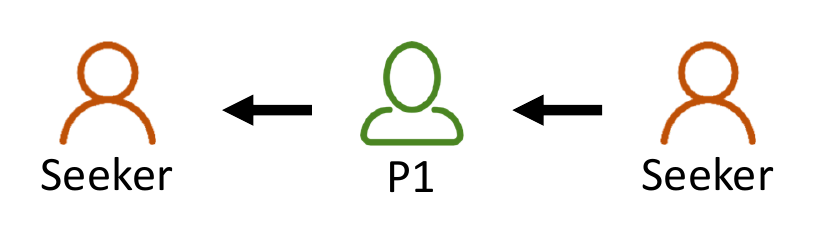} } 
\hfill
\subfloat[Mutual Discourse]{
	\label{subfig:fully-interactive}
	\includegraphics[width=0.75\columnwidth]{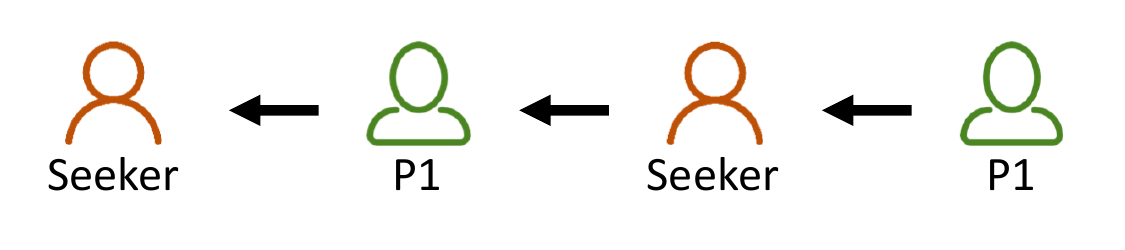} }
\caption{Three types of threads based on degree of interaction between seekers and peer-supporters based on models in communication theory~\cite{bretz1983media,williams1988research}.}
\label{fig:interactions}
\end{figure}

\subsection{Interaction-Based Indicators}
The second set of indicators capture how seekers \& peer-supporters interact with each other. The indicators are:

\xhdr{Time between Responses} The time difference between the consecutive posts in a thread. We differentiate between threads with small time between responses (\textit{Quick Threads}) and the threads with large time between responses (\textit{Slow Threads}). Again, instead of manually choosing a threshold we use a generative model to distinguish between the classes.

\xhdr{Degree of Interaction} To what extent do the seekers and peer-supporters interact in a thread? We identify three types of threads based on the degree of interaction motivated by communication theory (Figure~\ref{fig:interactions}). The first two types of threads are driven by a response from the seeker. The interactive communication theory by Bretz~\cite{bretz1983media} differentiates between two interaction mechanisms --- interactions in which the sender of a post gets a reply from the receiver but never responds back, and interactions in which the sender gets a reply from the receiver and also responds back. We extend these definitions to the context of threads on online support platforms and define \textit{Single Interaction Threads} and \textit{Repeated Seeker Interaction Threads}. A \textit{Single Interaction Thread} is one in which the seeker of the thread gets a reply from one or multiple peer-supporter(s) but never responds back (Figure~\ref{subfig:quasi-interactive}). And in a \textit{Repeated Seeker Interaction Thread}, the seeker responds back after a reply from peer-supporter(s) (Figure~\ref{subfig:interactive}). 
We further define a third type of thread, which corresponds to whether a peer-supporter, who had earlier posted on the thread, responds back after a response from the seeker (Figure~\ref{subfig:fully-interactive}). We call these threads \textit{Mutual Discourse}, a term coined in Williams et al.~\cite{williams1988research}, which relates to higher degrees of interaction between a sender and a receiver.


\section{Modeling Thread-Level Engagement}
\label{sec:model}

Given the aforementioned indicators of engagement, 
we aim to systematically discover patterns of engagement in threads. In order to achieve this, we perform the task of modeling thread-level engagement. We start by reasoning for the need of a computational model for discovering the engagement patterns. We then formally describe our modeling assumptions, followed by the generative process of the model and the parameter inference.

\label{subsec:model-motivation}
\xhqr{Why do we need a joint understanding of engagement indicators}
Analyzing only a single dimension of engagement for a thread is likely to be insufficient; it won't present a comprehensive view of the engagement dynamics of a thread. Consider the following two dimensions: number of peer-supporters and the degree of interaction (Figure~\ref{fig:interactions}). If we are looking at both of them individually, then we might miss out on cases where a seeker is having deep mutual discourse with a lot of peer-supporters; this simultaneous occurrence or interaction of both dimensions may have a stronger effect on engagement and the subsequent conversation outcome, than the occurrence of deep mutual discourse or a lot of peer-supporters, independently. Similarly, two threads with similar number of posts and peer-supporters can have very different engagement dynamics if one has long delays between messages from the peer supporters compared to the other.


\xhqr{Why do we need a model} Once the indicators are defined, the potential space of engagement patterns becomes obvious. Specifically, the engagement pattern $e$ of a thread would lie in the following space generated by the indicators:
\begin{align}
 e \in & \underbrace{\left\{ \text{\texttt{Short}, \texttt{Long}} \right\}}_{\text{What is the length?}}
 \times
 \underbrace{\left\{ \text{\texttt{Slow}, \texttt{Quick}} \right\}}_{\text{What is the Time between Responses?}} \nonumber \\
 & \times
 \underbrace{\left\{\text{\texttt{Isolated}, \texttt{Two-Party}, \texttt{Multi-Party}} \right\}}_{\text{How many Peer-Supporters?}} \nonumber \\
 & \times
  \left\{ \text{\texttt{Single Interaction}, \texttt{Repeated Seeker}} \right. \nonumber \\
 & \underbrace{\left. \text{ \texttt{Interaction, Mutual Discourse}} \right\}}_{\text{What is the degree of interaction between seeker and peer-supporter?}}
 \label{eq:taxonomy_space}
\end{align}

One might consider manually exploring engagement patterns across all combinations of the four engagement indicators. However, this approach would require extensive manual effort and time, and the need for making arbitrary decisions on thresholds (e.g. long vs. short threads), which would render the approach non-scalable, domain-dependent and subjective. Further, some indicators might be naturally correlated (e.g. thread length and number of peer supporters) in the data, obviating the need for defining certain classes of engagement patterns that are most likely empty or unnatural (e.g. long isolated threads).

\xhdr{Our Model} We propose a generative model over the set of engagement indicators described in Section~\ref{sec:indicators}. The model learns a set of clusters of threads based on a maximum likelihood objective (Equation 2). This type of modeling, as we shall see later, allows us to discover the distinct and interpretable engagement patterns of threads. We discuss the details of our model next.

\begin{figure}[t]
\centering
\subfloat[Length distribution of threads with $\delta_{i,j}>=100$; beta-fit]{
	\label{subfig:len-slow}
	\includegraphics[width=0.47\columnwidth]{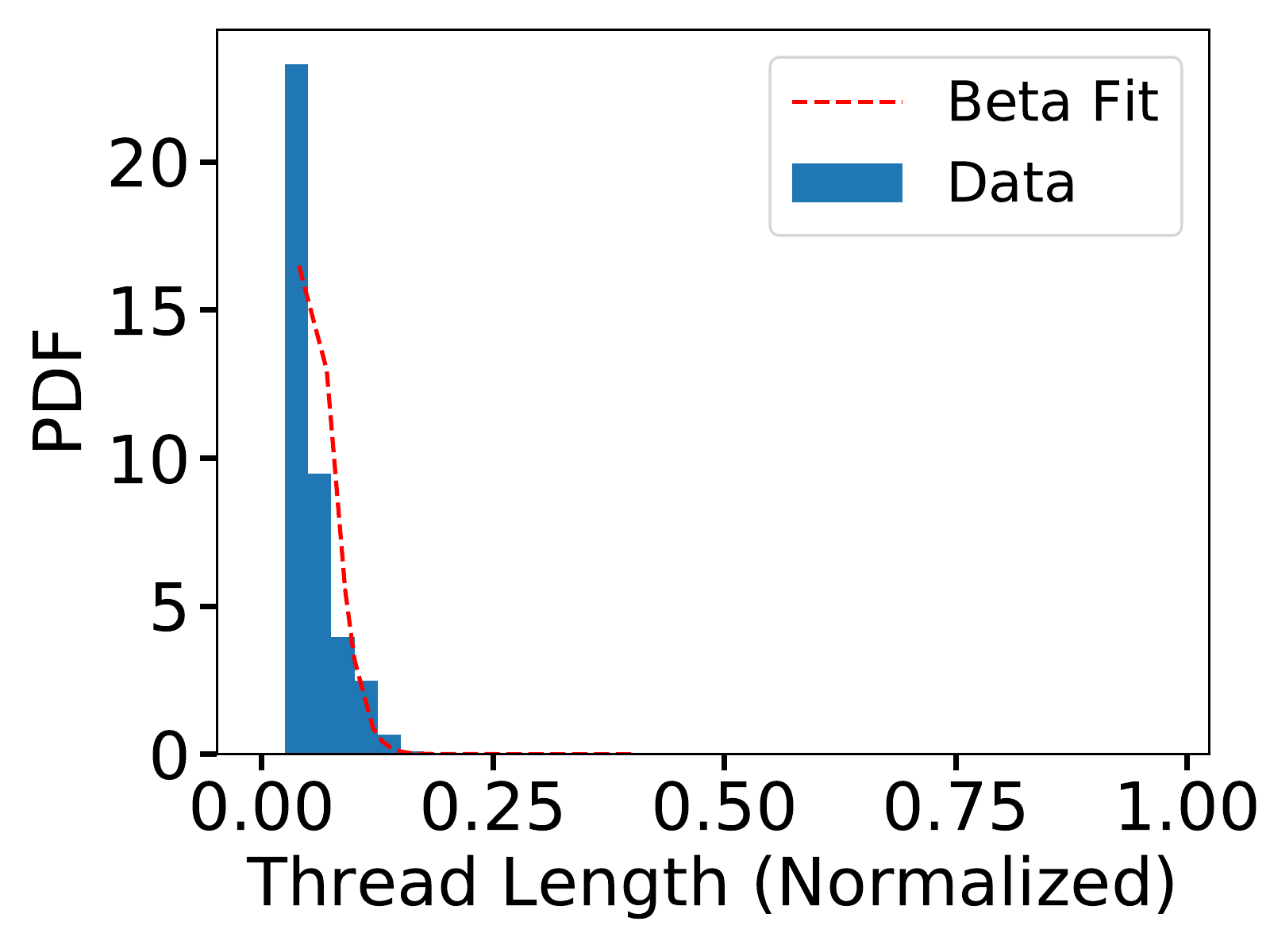} } 
\hfill
\subfloat[Length distribution of threads which are Multi-Party Mutual Discourse; beta fit]{
	\label{subfig:len-multi}
	\includegraphics[width=0.47\columnwidth]{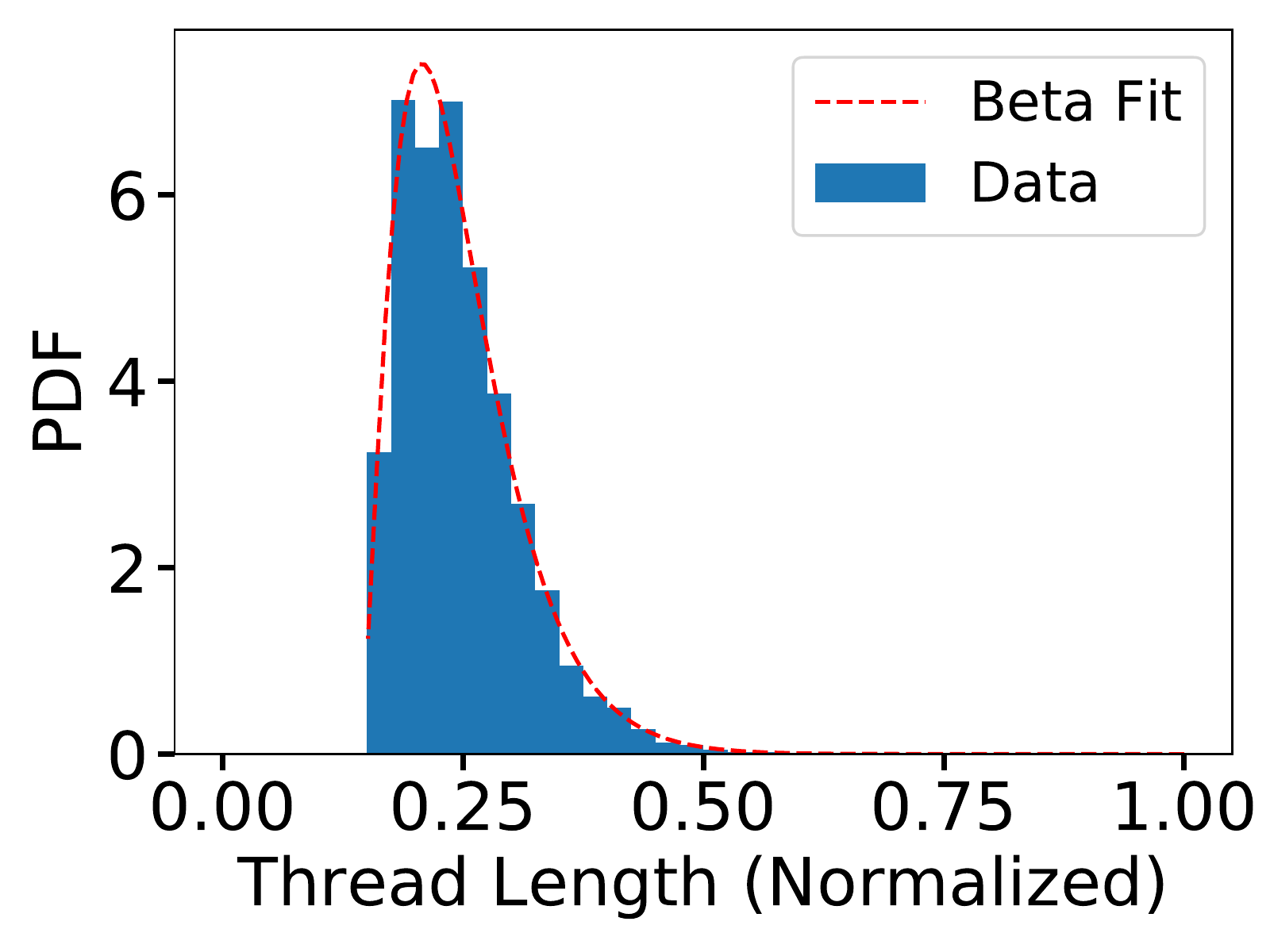} } 
\caption{Empirical validation of modeling assumptions.}
\label{fig:model-assumptions}
\end{figure}
\subsection{Modeling Assumptions}
We assume that every cluster has {its} own joint distribution over the engagement indicators of threads; each joint distribution describes a different thread-level engagement pattern. We further assume that each thread is generated from a single engagement cluster. This is helpful in efficient learning of our model parameters primarily due to limited signals available in each thread~\cite{yin2014dirichlet}.

Let $\mathcal{T}$ be the set of threads. A thread $\mathbf{T}_{i} \in \mathcal{T}$ consists of an initial post $p_{i,0}$ by the seeker and a set of $k-1$ replies $p_{i,1}, p_{i,2}, ... , p_{i,k-1}$ having a total thread length of $k$\footnote{We combine two consecutive posts by the same user as a pre-processing step. Note that this may result in loss of temporal effects, which is not in the scope of this study.}. Here, we represent each reply in the thread with a tuple $p_{i,j} = (u_{i,j}, r_{i,j}, \delta_{i,j})$ where $u_{i,j}$ is the user of the post, $\delta_{i,j}$ is the time elapsed since the last post and $r_{i,j}$ is the role of the user based on the interaction dynamics local to the thread (indicative of first peer-supporter, re-entry of an existing peer-supporter, a new peer-supporter, and seeker's response). We define a set $\mathcal{R}$ of 4 user roles: \textbf{(a)} \textbf{First Peer-Supporter:} user of the first reply of the thread, i.e., $j = 1$; \textbf{(b)} \textbf{New Peer-Supporter}: $u_{i,j}$ is new to the thread but not the first peer-supporter, i.e., $\forall k < j \; u_{i,k}$ $\neq$ $u_{i,j}, j \neq 1$; \textbf{(c)} \textbf{Existing Peer-Supporter}: $u_{i,j}$ is a peer-supporter who has interacted with the thread before, i.e., $\exists$ $k<j$: $u_{i,k} = u_{i,j}$ and $u_{i,k} \neq u_{i,0}$; \textbf{(d) Seeker}: $u_{i,j}$ is the seeker, i.e., $u_{i,j} = u_{i,0}$. This categorization of user types based on interaction helps us in accounting for both the number of peer-supporters and the degree of interaction between the seeker and the peer-supporter. 
We intentionally distinguish between first peer-supporter and new peer-supporter as this allows us to easily differentiate between \texttt{Two-Party} threads and \texttt{Multi-Party} threads; a thread would be \texttt{Two-Party} if there is no new peer-supporter. 
Moreover, the use of seeker and existing peer-supporter is helpful in differentiating between \texttt{Single Interaction}, \texttt{Repeated Seeker Interaction}, and \texttt{Mutual Discourse} threads. 

\xhdr{Parametric Assumptions} 
We assume that the distributions of thread lengths and time between responses can be well-approximated through Beta distributions. 
This assumption is based in both theoretical and empirical findings.
Theoretically, beta distributions can approximate power laws as well as family of exponential  distributions that emerge in natural systems that follow some kind of preferential attachment law~\cite{peruani2007emergence}.

Empirically, thread lengths in online forums usually follow a power-law distribution~\cite{yu2010analyzing}. Further, as shown in Figure~\ref{fig:model-assumptions},  the length\footnote{We use min-max scaling for transforming the lengths and time between responses to $[0, 1]$ interval, which then can be modeled by beta-distribution.} distribution of threads on \textsc{TalkLife}---(i) which are potentially slow ($\delta_{i,j} >= 100$) (Figure~\ref{subfig:len-slow}), and (ii) in which  the seeker has \texttt{Mutual Discourse} with multiple peer-supporters (\texttt{Multi-Party}) (Figure~\ref{subfig:len-multi}) -- are both well-approximated by Beta distributions.

We make similar observations for the potential $\delta_{i,j}$ clusters. We make use of these observations in our model; we assume that the lengths and the time between responses within each cluster are generated from a Beta distribution.

Moreover, user roles in a thread are assumed to be categorical distributions over the set $\mathcal{R}$ consisting of the 4 types of roles --- First Peer-Supporter, New Peer-Supporter, Existing Peer-Supporter and Seeker, where the distributions themselves are Dirichlet distributed (similar to model assumptions of Latent Dirichlet Allocation~\cite{blei2003latent}).

\subsection{Generative Process}
Let $\mathcal{E}$ be the set of engagement clusters. A thread $\mathbf{T}_{i} \in \mathcal{T}$ is generated as described in Algorithm~\ref{alg:dup}. The engagement distribution $\theta_{\mathcal{E}}$ is drawn from a Dirichlet distribution with prior $\alpha_{\mathcal{E}}$ (Line 1). Likewise, for every engagement cluster $e$, user-role distributions $\phi_{e}^{\mathcal{R}}$ are drawn from a Dirichlet distribution with prior $\alpha_{e}^{\mathcal{R}}$ (Line 2-4). For every thread $\mathbf{T}_{i}$, first an engagement cluster $e$ is chosen from the engagement distribution $\theta_{\mathcal{E}}$ (Line 6). The thread length $k$ of $\mathbf{T}_{i}$ is sampled from the beta distribution of this engagement cluster $e$ parameterized by alpha $\alpha_{e}^{\mathcal{K}}$ and beta $\beta_{e}^{\mathcal{K}}$ (Line 7). Next, the engagement cluster generates the set of replies (Line 8-11). The $j$-th reply consists of local user roles $r_{i,j}$ and time deltas $\delta_{i,j}$. Each $r_{i,j}$ is sampled from the categorical user-role distribution $\phi_{e}^{\mathcal{R}}$ (Line 9) and each $\delta_{i,j}$ is sampled from the beta distribution parameterized by alpha $\alpha_{e}^{\delta}$ and beta $\beta_{e}^{\delta}$ (Line 10). The likelihood of generating thread $T_i$ from an engagement cluster $e$ is given by:
\resizebox{.93\columnwidth}{!}{
  \begin{minipage}{\columnwidth}
\begin{align*}
\label{eq:likelihood}
    p(\mathbf{T}_{i} | e) & \propto\frac{n_{e} + \alpha_{\mathcal{E}}}{|\mathcal{T}| + |\mathcal{E}| * \alpha_{\mathcal{E}}} * \frac{k^{\alpha_{e}^{\mathcal{K}}-1} (1-k)^{\beta_{e}^{\mathcal{K}}-1}}{B(\alpha_{e}^{\mathcal{K}}, \beta_{e}^{\mathcal{K}})} \nonumber \\
    & *  \displaystyle\prod_{p_{i,0}, p_{i,1}, ... , p_{i,k-1}} \left(\phi_{e}^{\mathcal{R}} (r_{i,j}) * \frac{\delta_{i,j}^{\alpha_{e}^{\mathcal{\delta}}-1} (1-\delta_{i,j})^{\beta_{e}^{\mathcal{\delta}}-1}}{B(\alpha_{e}^{\mathcal{\delta}}, \beta_{e}^{\mathcal{\delta}})} \right) 
\end{align*}
\end{minipage}
}
\hfill (2)

\stepcounter{equation}
\noindent where $n_e$ is the number of threads in the engagement cluster $e$.

Our learning objective for deriving the desired clusters is to maximize this likelihood given a dataset of threads. 
Note that we do not make use of the \texttt{Isolated} threads while learning the model; these threads are separately identified and integrated with the inferred patterns (Section~\ref{sec:taxonomy}).

\vspace{5pt}
\subsection{Parameter Inference}
\vspace{5pt}
We use a Gibbs-sampling approach for inferring the Dirichlet distribution. The likelihood of generating a role $r_{i,j}$ from engagement cluster $e$ is given by:
\begin{align}
    \phi_{e}^{\mathcal{R}} (r_{i,j}) & = \frac{n_{e}^{(r_{i,j})} + \alpha_{\mathcal{R}}}{n_{e}^{(.)} + |\mathcal{R}|* \alpha_{\mathcal{R}}}
\end{align}
where $n_{e}^{(r_{i,j})}$ is the number of times role $r_{i,j}$ has been assigned to cluster $e$, $n_{e}^{(.)}$ is the marginal count over all roles in $\mathcal{R}$. We use method of moments for inferring the Beta distribution parameters --- $\alpha_{e}^{\mathcal{K}}$, $\beta_{e}^{\mathcal{K}}$, $\alpha_{e}^{\delta}$, $\beta_{e}^{\delta}$~\cite{wang2006topics}. We initialize the two Dirichlet priors using a commonly used strategy in LDA-based models ($\alpha_{X} = 50/|X|, X = {\mathcal{E}, \mathcal{R}}$)~\cite{diao2012finding}. We optimize on the number of clusters using the popular Elbow method. For both \textsc{TalkLife} and \textsc{Reddit}, we choose the number of clusters as 20 which also gives us the most diverse and interpretable clusters.

\vspace{5pt}
\section{Inferred Engagement Patterns}
\vspace{5pt}
\label{sec:taxonomy}

\begin{algorithm}[t]
 \caption{Generative process of our engagement model}
 \begin{algorithmic}[1]
 \State Draw engagement distribution $\theta_{\mathcal{E}}$ $\sim$ $Dir(\alpha_{\mathcal{E}})$
  \For{each engagement cluster $e \in \mathcal{E}$}
  \State Draw user-role distribution $\phi_{e}^{\mathcal{R}}$ $\sim$ $Dir(\alpha_{\mathcal{R}})$
  \EndFor
  \For{each thread $\mathbf{T}_{i} \in \mathcal{T}$}
  \State Draw an engagement cluster $e$ $\sim$ $\theta_{\mathcal{E}}$
  \State Draw the thread length $k$ $\sim$ $Beta(\alpha_{e}^{\mathcal{K}}, \beta_{e}^{\mathcal{K}})$
  \For{each reply post $p_{ij}$ $\in$ $\mathbf{T}_{i}$}
  \State Draw the user role $r_{ij}$ $\sim$ $Multi(\phi_{e}^{\mathcal{R}})$
  \State Draw the time to reply $\delta_{ij}$ $\sim$ $Beta(\alpha_{e}^{\mathcal{\delta}}, \beta_{e}^{\mathcal{\delta}})$
  \EndFor
  \EndFor
 \end{algorithmic}
 \label{alg:dup}
\end{algorithm}

Using the engagement clusters learned by our generative model, we infer the predominant set of engagement patterns of threads on \textsc{TalkLife} and \textsc{Reddit}. We analyze the distributions which the learned clusters have over the three dimensions of a thread (length, time delta, user roles) --- 
$Beta(\alpha_{e}^{\mathcal{K}}$,  $\beta_{e}^{\mathcal{K}})$,  
$Beta(\alpha_{e}^{\mathcal{\delta}}$,
$\beta_{e}^{\mathcal{\delta}})$, and
$\phi_{e}^{\mathcal{R}}$. We do this by simultaneously considering the space over engagement indicators defined in Equation~\ref{eq:taxonomy_space}. The clusters are manually analyzed and coded by multiple authors in a top-down approach, after which we derive the following \textit{hierarchically organized engagement patterns} (fraction of threads following each pattern is shown in brackets -- \textsc{TalkLife} \& \textsc{Reddit} respectively):

\begin{enumerate}[label=$\bullet$]

    \item \textbf{Isolated (32.43\% \& 27.53\%)}
    
    \item \textbf{Single Interaction (30.57\% \& 7.64\%):} 
    
    \begin{enumerate}[label=$\circ$]

        \item Two-Party (20.30\% \& 0.08\%):
        
        \begin{enumerate}[label=(\roman*), leftmargin=2\parindent]
            \item Short Slow Two-Party SI (20.30\% \& 0.08\%)
        \end{enumerate}
        
        \item Multi-Party (10.27\% \& 7.56\%):
        \begin{enumerate}[label=(\roman*), leftmargin=2\parindent]
        \setcounter{enumiii}{1}
            \item Short Slow Multi-Party SI (10.27\% \& 7.56\%)
        \end{enumerate}
        
    \end{enumerate}
    
    \item \textbf{Repeated Seeker Interaction (18.6\% \& 21.4\%):}
    
    \begin{enumerate}[label=$\circ$]

        \item Two-Party (4.25\% \& 5.58\%):

        \begin{enumerate}[label=(\roman*), leftmargin=2\parindent]
        \setcounter{enumiii}{2}
            \item Short Slow Two-Party RSI (3.39\% \& 3.99\%)
            \item Short Quick Two-Party RSI (0.86\% \& 1.59\%)
        \end{enumerate}
        
        \item Multi-Party (14.35\% \& 15.82\%):
        \begin{enumerate}[label=(\roman*), leftmargin=2\parindent]
        \setcounter{enumiii}{4}
            \item Short Slow Multi-Party RSI (1.10\% \& 12.96\%)
            \item Short Quick Multi-Party RSI (13.25\% \& 2.86\%)
        \end{enumerate}
        
    \end{enumerate}
    
    \item \textbf{Mutual Discourse (18.4\% \& 43.43\%):}
    
    \begin{enumerate}[label=$\circ$]

        \item Two-Party (8.86\% \& 22.08\%):

        \begin{enumerate}[label=(\roman*), leftmargin=2\parindent]
        \setcounter{enumiii}{6}
            \item Short Quick Two-Party MD (8.11\% \& 21.99\%)
            \item Long Quick Two-Party MD (0.75\% \& 0.09\%)
        \end{enumerate}
        
        \item Multi-Party (9.54\% \& 21.35\%):
        \begin{enumerate}[label=(\roman*), leftmargin=2\parindent]
        \setcounter{enumiii}{8}
            \item Short Quick Multi-Party MD (6.17\% \& 17.33\%)
            \item Long Quick Multi-Party MD (3.37\% \& 4.02\%)
        \end{enumerate}
        
    \end{enumerate}

\end{enumerate}

where SI: Single Interaction; RSI: Repeated Seeker Interaction and MD: Mutual Discourse. The engagement patterns are named based on the most likely or dominant set of engagement indicators. We qualitatively evaluate the inferred patterns as described next.

\begin{figure}[b]
\centering
\vspace{-10pt}
\includegraphics[width=0.9\columnwidth]{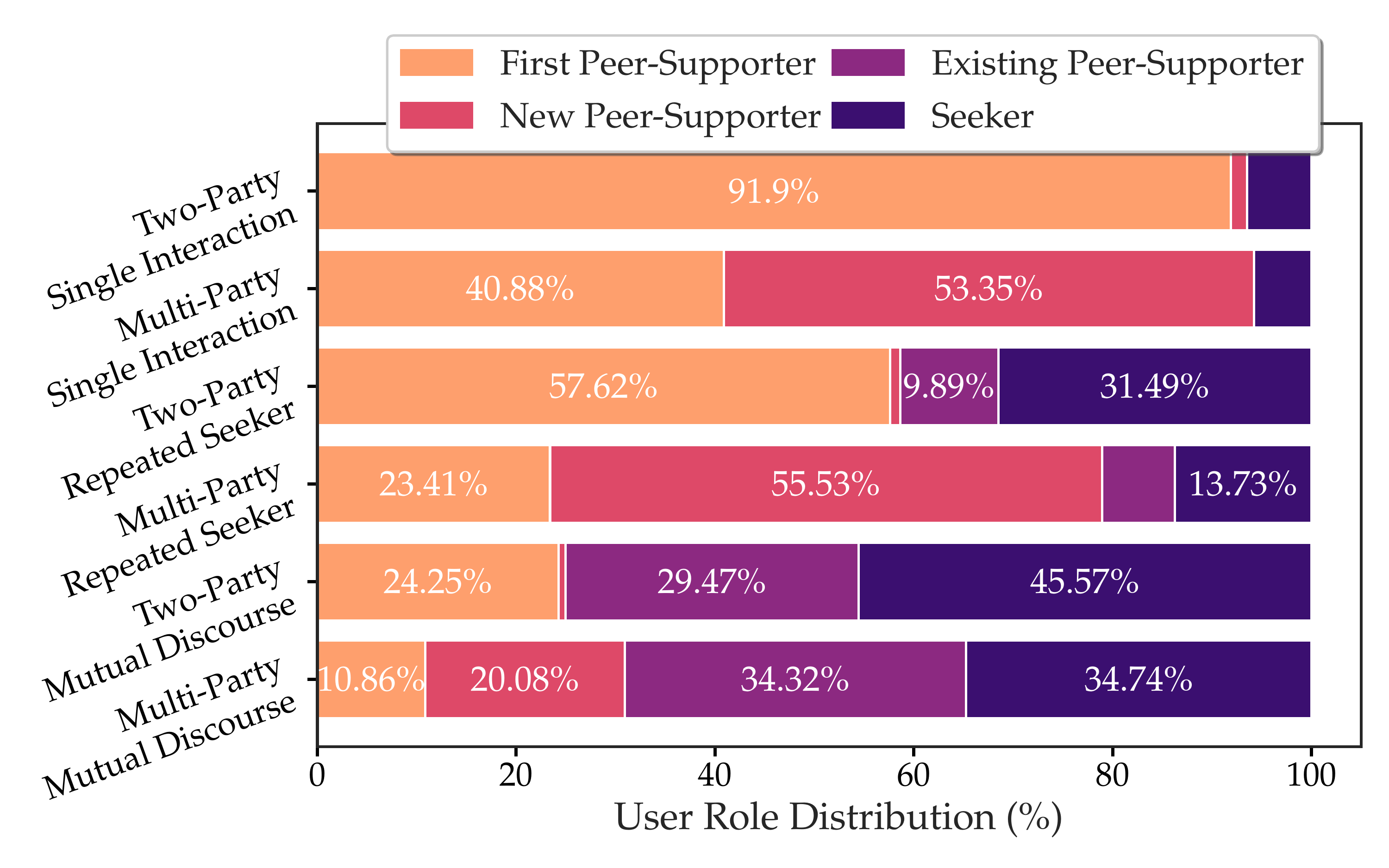}
\caption{Engagement patterns and user roles.}
\label{fig:user-role-distr}
\end{figure}

\subsection{Qualitative evaluation of inferred  patterns}
We present a qualitative evaluation of the distributions of the inferred engagement patterns over user roles, thread length and time between responses.

\xhdr{User roles}
Figure~\ref{fig:user-role-distr} shows the distribution of engagement patterns over user roles (marginalizing the patterns along length and time between responses). The \texttt{Two-Party} patterns contain high percentage of peer-supporters who are first to the threads (first peer-supporters) and low percentage of peer-supporters who are new (new peer-supporter); this is indicative of presence of only one peer-supporter (the first one) in the thread, hence \texttt{Two-Party} (the other party being the seeker). On the other hand, \texttt{Multi-Party} patterns have higher percentage of peer-supporters who are new, indicative of multiple peer-supporters in the thread. The \texttt{Single Interaction} patterns have low percentage of seekers and existing peer-supporters; these patterns rarely get responses from seekers and existing peer-supporters. Finally, the \texttt{Mutual Discourse} threads have high seeker and existing peer-supporter percentages, indicative of multiple interaction between the seeker and peer-supporter(s) of the thread.

\xhdr{Thread lengths \& time between responses}
Further, we analyze the distributions of the inferred \texttt{Short}, \texttt{Long}, \texttt{Slow} \& \texttt{Quick} threads for both \textsc{TalkLife} \& \textsc{Reddit}. On \textsc{TalkLife}, \texttt{Short} threads have a mean length of 3.9 and a median length of 3. Whereas \texttt{Long} threads have a mean length of 13.5 and a median length of 10. \texttt{Slow} threads have a median time to reply of 7 minutes. \texttt{Quick} threads have a median time to reply  of 1 minutes.

On \textsc{Reddit}, \texttt{Short} threads have a mean length of 3.33 and a median length of 3. Whereas \texttt{Long} threads have a mean length of 23.95 and a median of 19. \texttt{Slow} threads have a median time to reply of 75 minutes. \texttt{Quick} threads have a median time of 16 minutes.

\begin{figure}
\centering
\subfloat[Thread Length]{
	\label{subfig:seeker-retention-len}
	\includegraphics[width=0.47\columnwidth]{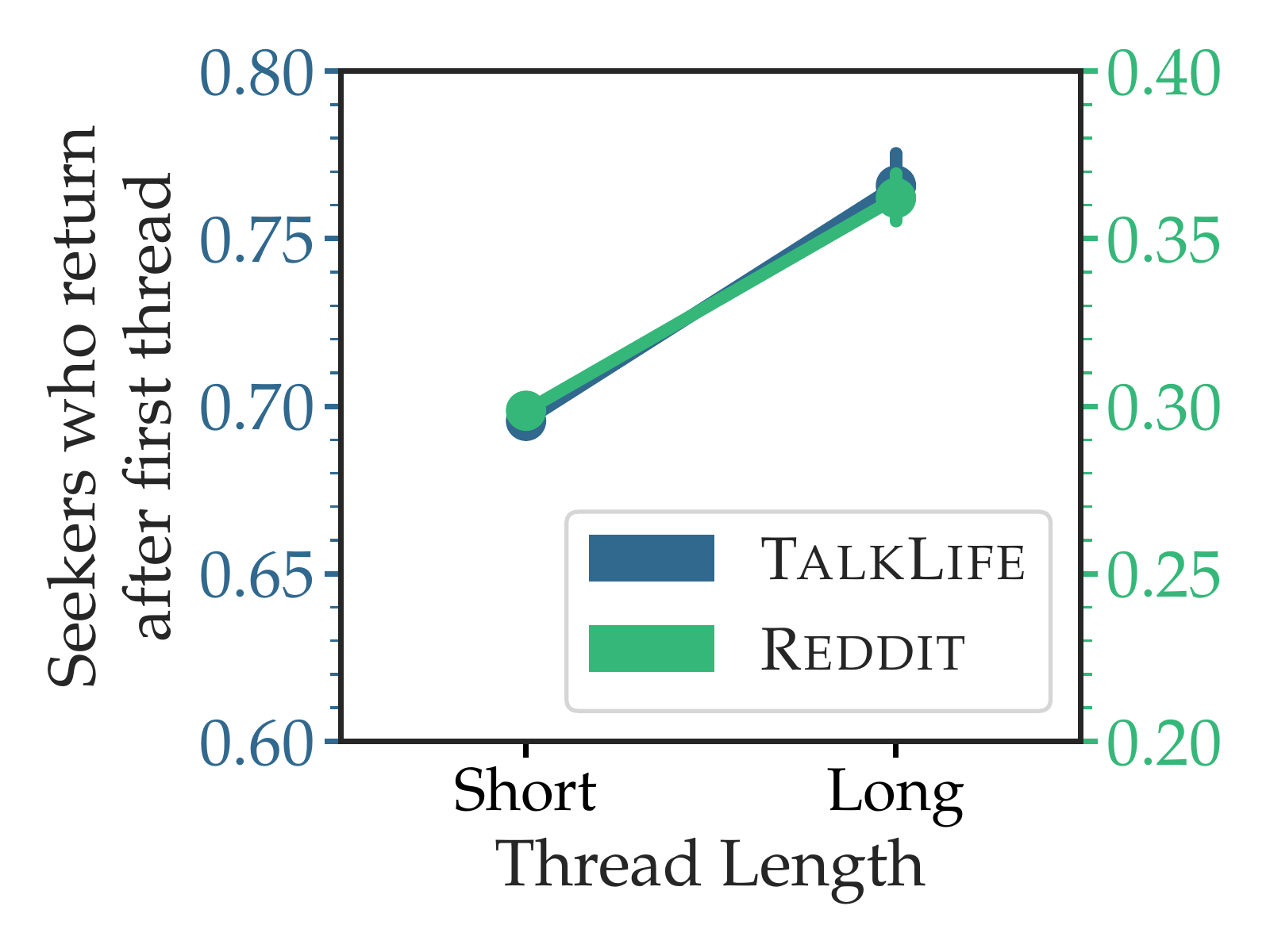} } 
\hfill
\subfloat[Time between Responses]{
	\label{subfig:seeker-retention-time}
	\includegraphics[width=0.47\columnwidth]{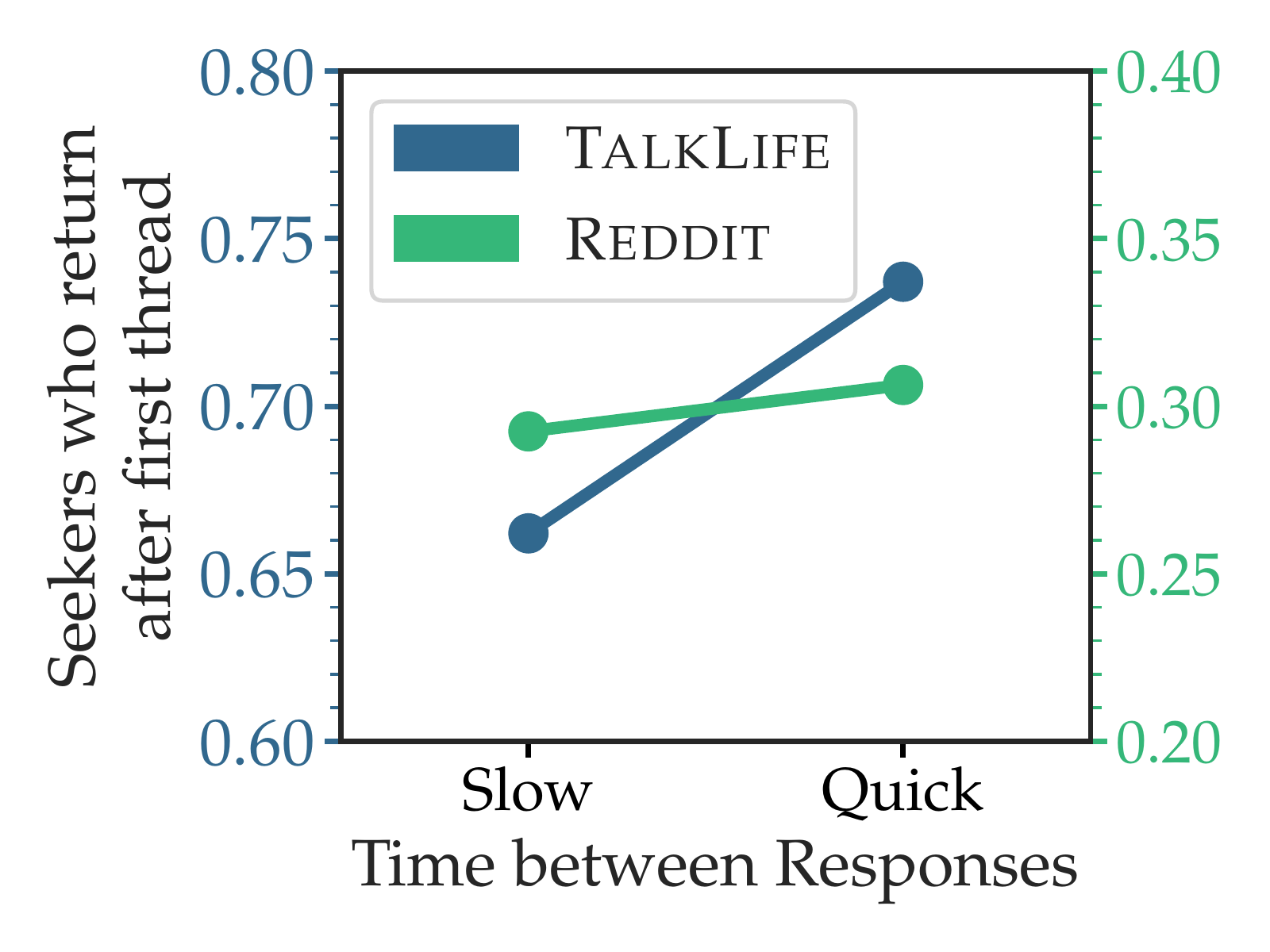} } 
\hfill
\subfloat[Number of Peer-Supporters]{
	\label{subfig:seeker-retention-peer-supp}
	\includegraphics[width=0.47\columnwidth]{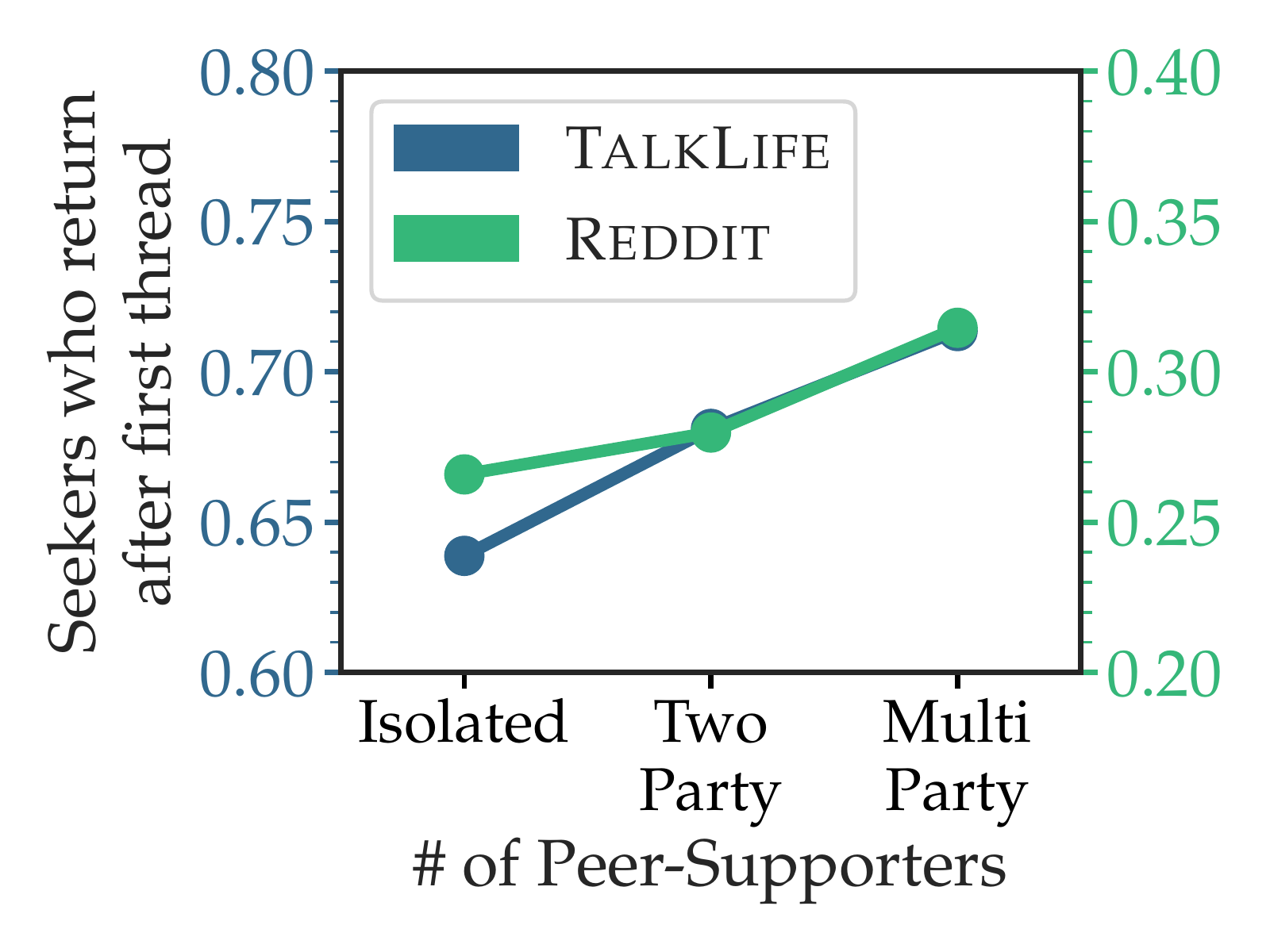} } 
\hfill
\subfloat[Degree of Interaction]{
	\label{subfig:seeker-retention-interaction}
	\includegraphics[width=0.47\columnwidth]{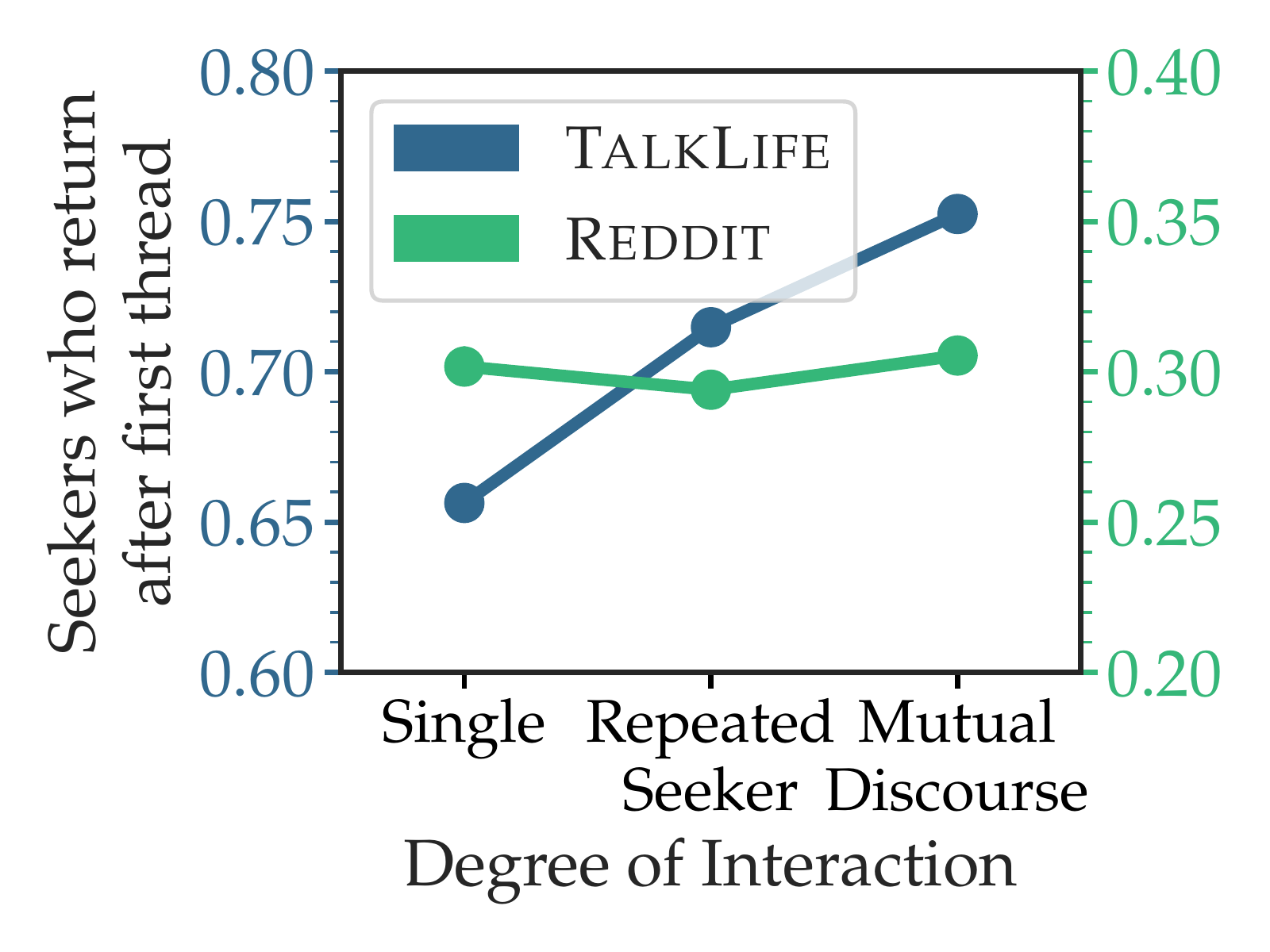} } 
\caption{Fraction of seekers who return after their first thread across four engagement indicators. Seekers are more likely to return after higher degrees of engagement. Note that, in this paper, we are only interested in variation of retention likelihood with degrees of engagement and not the absolute values. Error bars throughout the paper are bootstrapped 95\% confidence intervals.}
\label{fig:seeker-retention}
\end{figure}

\begin{figure*}
\centering
\subfloat[Single Interaction Patterns]{
	\label{subfig:seeker-retention-joint-si}
	\includegraphics[width=0.47\columnwidth]{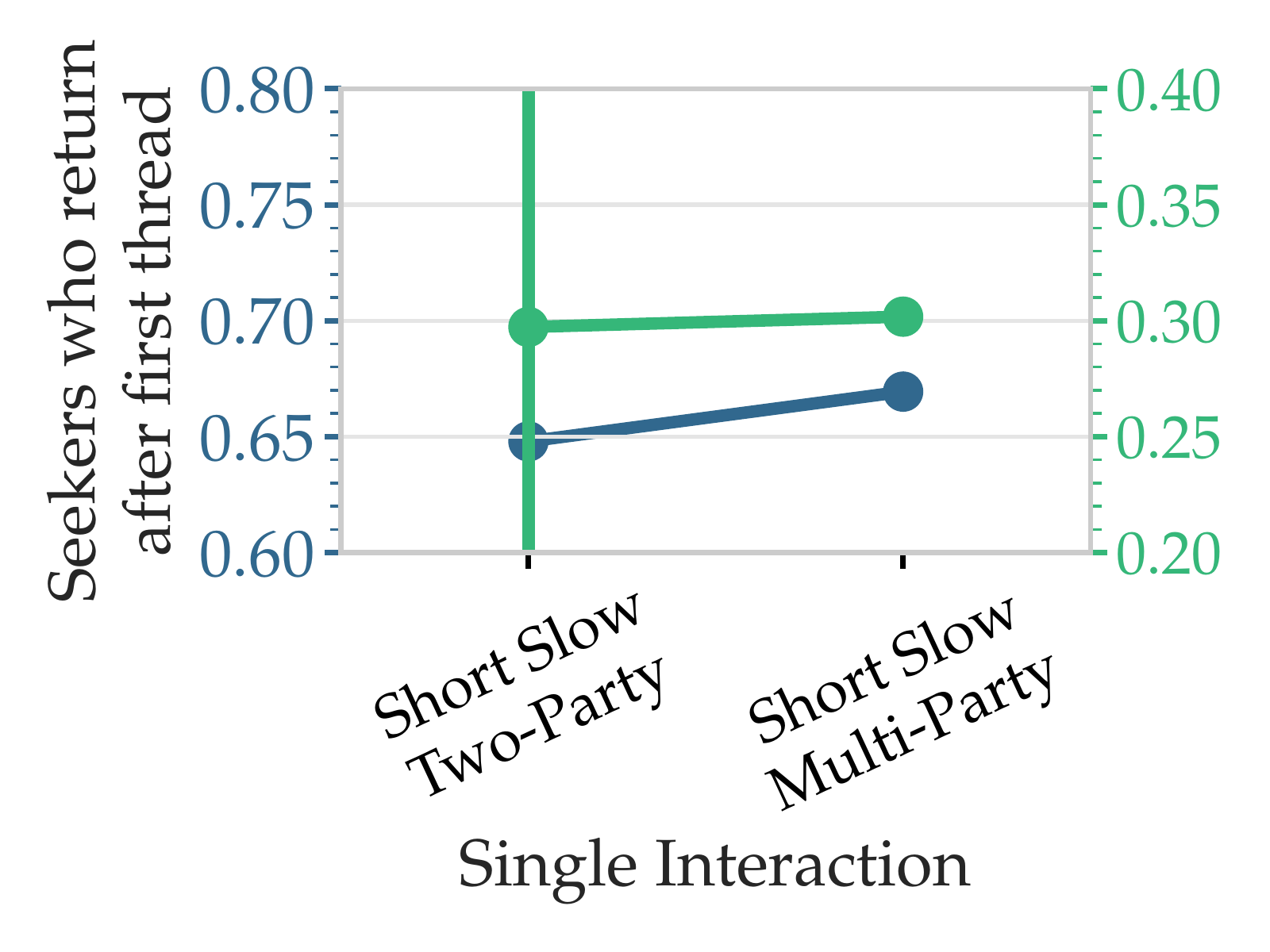} } 
\hfill
\subfloat[Repeated Seeker Interaction Patterns]{
	\label{subfig:seeker-retention-joint-rsi}
	\includegraphics[width=0.7\columnwidth]{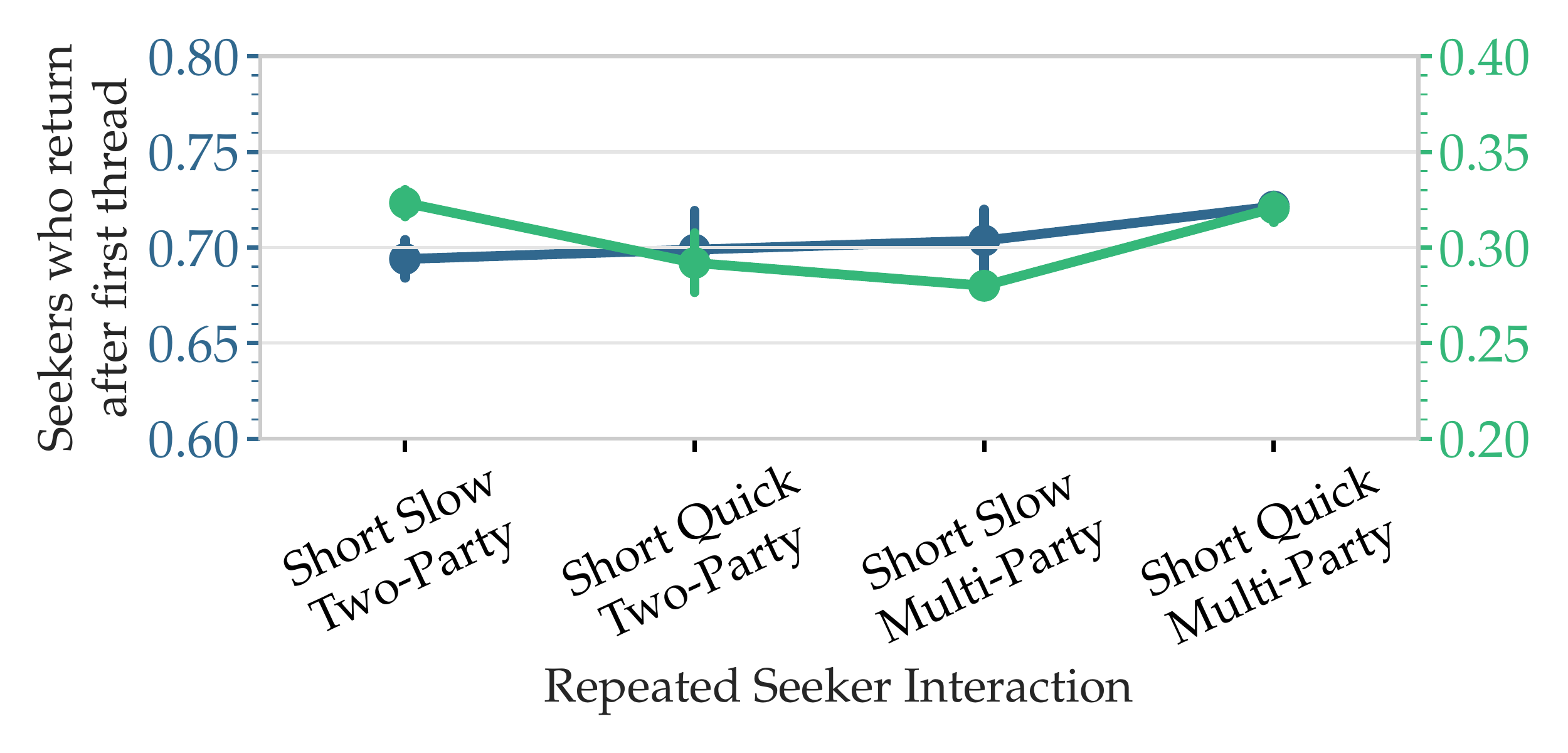} } 
\hfill
\subfloat[Mutual Discourse Patterns]{
	\label{subfig:seeker-retention-joint-md}
	\includegraphics[width=0.7\columnwidth]{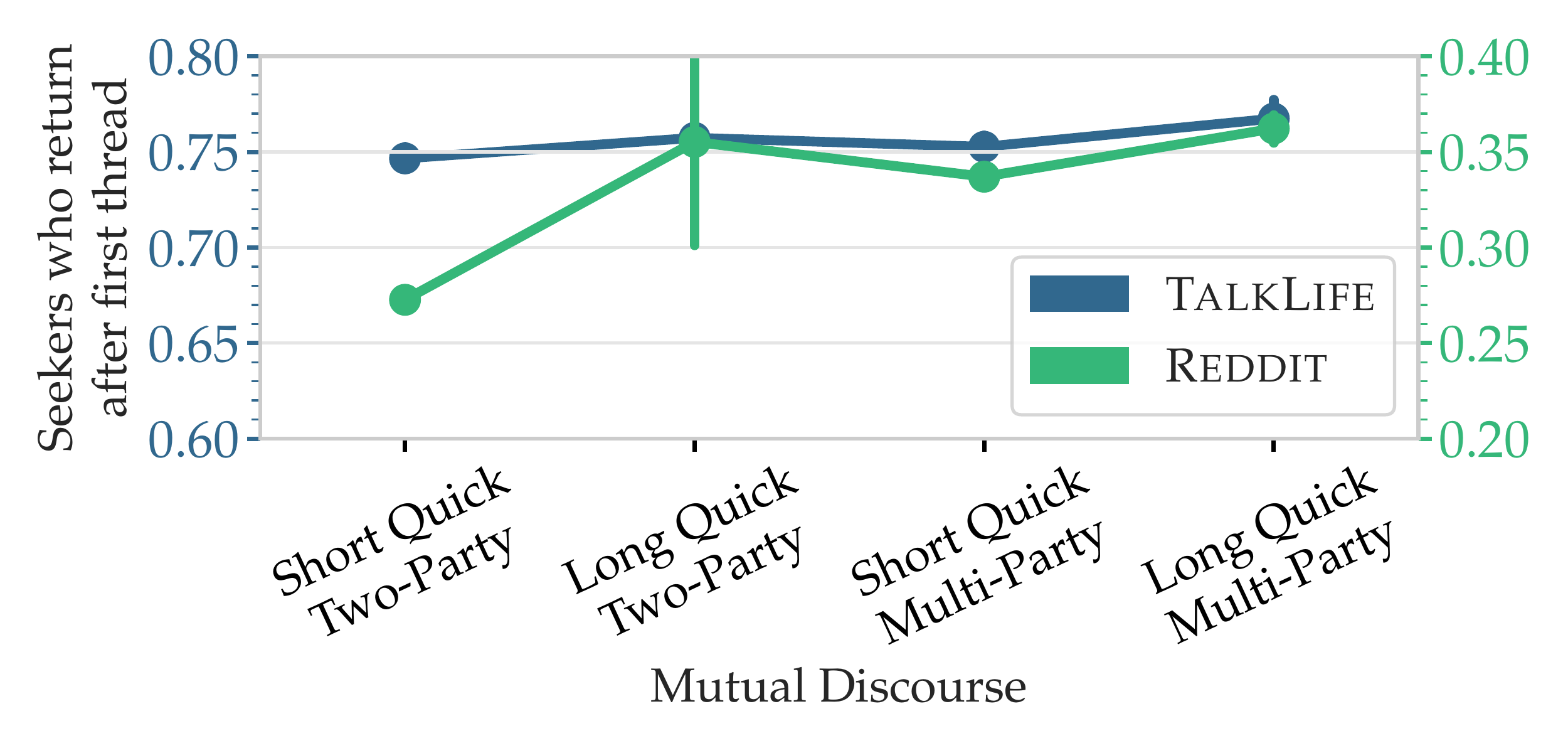} } 
\caption{Seeker retention and joint engagement patterns. Degree of Interaction is key to seeker retention; \texttt{Mutual Discourse} is more important for seeker retention than  \texttt{Repeated Seeker Interaction} or \texttt{Single Interaction}. Within each interaction degree there is very limited variation between engagement patterns.}
\label{fig:seeker-retention-joint}
\end{figure*}


\section{Implications of Engagement Patterns}
\label{sec:insights}

We next investigate the implications of our framework of engagement patterns and demonstrate ways in which it can guide online support platform evaluations and assessments. 

We first exploit our framework for comparing the functioning of \textsc{TalkLife} and \textsc{Reddit} (Section~\ref{subsec:compare}). Next, we look at the correlations these engagement patterns have with the retention of seekers (Section~\ref{subsec:seeker-retention}) and peer-supporters (Section~\ref{subsec:ps-retention}) on \textsc{TalkLife} and \textsc{Reddit}. Before we conclude, we investigate the factors which potentially lead to \texttt{Mutual Discourse} between seekers and peer-supporters in a thread (Section~\ref{subsec:mutual-discourse}). 

\subsection{Comparative assessment of \textsc{TalkLife} \& \textsc{Reddit}}
\label{subsec:compare}

The patterns inferred in Section~\ref{sec:taxonomy} exhibit an interesting contrast between \textsc{TalkLife} and \textsc{Reddit}. Even though the amount of \texttt{Isolated} threads on the two platforms is comparable, we observe \textsc{Reddit} to have very few \texttt{Single Interaction} threads involving one peer-supporter (\texttt{Two-Party}; 0.08\%). This suggests that, on \textsc{Reddit}, after an interaction happens on a thread posted by a seeker, it either attracts other users or it spawns a future interaction with the seeker. Also, a much larger fraction of threads on \textsc{Reddit} result in a \texttt{Mutual Discourse} as compared to \textsc{TalkLife} (43.43\% vs. 18.4\%). Both these differences might be a result of the sub-community nature of \textsc{Reddit} where users can subscribe to the topics they care about and the topics in which they can provide effective peer-support. For example, a peer-supporter who has dealt with post traumatic stress disorder (PTSD) in the past might be looking for supporting users with PTSD; this can be achieved on the \textit{r/ptsd} subreddit. On \textsc{TalkLife}, however, that peer-supporter will need to manually \textit{follow} users having PTSD which may not be intuitive. This indicates that having an organized structure of the threads and users is helpful in making online platforms more engaging. Future work should investigate the relationship of engagement with community structure dynamics.

\subsection{Seeker Retention on support platforms}
\label{subsec:seeker-retention}
Seeker retention, particularly during the initial days of the seeker on the platform, is important for receiving proper support from the peers. We define retention as returning back to the platform after the first thread of interaction by writing another post or response in a different thread (on any of the mental health subreddits in case of \textsc{Reddit}). We look into the fraction of seekers who return back when their first threads have a certain engagement pattern.

\xhdr{Seeker retention increases with higher degrees of engagement} We examine how seeker retention varies with different degrees of engagement. Figure~\ref{fig:seeker-retention} shows the variation of fraction of seekers who return to the platform after their first threads across the four individual engagement indicators. We observe that seekers are more likely to return after higher degrees of engagement in their first thread; seeker retention is least likely after an \texttt{Isolated} first thread (0.64 \& 0.26 respectively for \textsc{TalkLife} \& \textsc{Reddit}); \texttt{Long} threads have more seeker retention likelihood than \texttt{Short} threads (0.77 vs. 0.70; 0.36 vs. 0.30); \texttt{Multi-Party} more than \texttt{Two-Party} (0.71 vs. 0.68; 0.31 vs. 0.28). Furthermore, on \textsc{TalkLife}, \texttt{Quick} threads have higher seeker retention likelihood than \texttt{Slow} (0.74 vs. 0.66); \texttt{Mutual Discourse} has higher likelihood than \texttt{Repeated Seeker Interaction} and \texttt{Single Interaction} threads (0.75 vs. 0.71 vs. 0.66). This, however, is in contrast with \textsc{Reddit}, where there is very little variation in seeker retention rates along the indicators of time between responses (0.29 vs. 0.30) and degree of interaction (0.31 vs. 0.29 vs. 0.31). 

\xhdr{\texttt{Mutual Discourse} is more important for seeker retention independent of other engagement indicators} We next check if the joint presence of the engagement indicators have a role to play on seeker retention. For this, we analyze the variation of seeker retention likelihood with joint engagement patterns. We find that the variation in the likelihood is largely between the three degrees of interaction; the variation is low among the other indicators within a specific degree of interaction (Figure~\ref{fig:seeker-retention-joint}). \texttt{Single Interaction} threads have the least likelihood followed by \texttt{Repeated Seeker Interaction}; \texttt{Mutual Discourse} have the highest retention likelihood. The variation is limited within each degree for \textsc{TalkLife}; a \texttt{Mutual Discourse} is expected to have the highest seeker retention likelihood independent of whether it is \texttt{Long} or \texttt{Short}, \texttt{Two-Party} or \texttt{Multi-Party}. 

To understand how these variations inform us better engagement mechanisms on \textsc{TalkLife}, consider a situation where a seeker starts a thread. The likelihood of her returning to \textsc{TalkLife} if she gets no response is 64\%. Receiving a single response from a peer-supporter  increases the retention likelihood to 65\%. However, if the seeker replies back after the response, likelihood jumps to 69\%. At this point, even if a few more peer-supporters post on the thread, the likelihood hovers around 70\%. This likelihood gets a major boost and jumps to 75\% if one of the existing peer-supporters replies again leading to a \texttt{Mutual Discourse}. This is how important \texttt{Mutual Discourse} is to \textsc{TalkLife}, with a gain of 10\% over \texttt{Single Interaction} threads and a gain of 5\%  over \texttt{Repeated Seeker Interaction} threads in terms of seeker retention likelihood.


This provides quantitative evidence that simply connecting users on the platform or trying to have each post get a response may not result in optimal outcomes. Instead, truly mutual interactions were associated with high seeker retention rates. Future work should investigate how to effectively engage seekers and peer-supporters in mutual discourse.

\textsc{Reddit}, surprisingly, has a low seeker retention likelihood for mutual discourses which are \texttt{Short}, \texttt{Quick}, and \texttt{Two-Party}. This hints towards platform-specific nuances on \textsc{Reddit}. These are instances of seekers having solitary mutual, back-and-forth conversations with a peer-supporter. A lot of them might have been \textit{throwaway} accounts on \textsc{Reddit} or the seeker involved might have moved to a different more-suitable subreddit.

\subsection{Peer-Supporter Retention}
\label{subsec:ps-retention}

Unsurprisingly, peer-supporters are key to a peer-to-peer support platform. Increasing their retention and ensuring that they, as a whole, are providing adequate support on the platform is critical for successful social support. We next investigate retention of peer-supporters and explore the engagement patterns which have an increased likelihood of peer-supporter retention. Similar to our seeker retention analysis, we define retention of a peer-supporter as returning back to the platform after the first thread of interaction (in a new thread). We inspect the correlation of engagement patterns of the threads with peer-support retention likelihood. 


\begin{figure}
\centering
\subfloat[\textsc{TalkLife}]{
	\label{subfig:ps-retention-joint-talklife}
	\includegraphics[width=0.47\columnwidth]{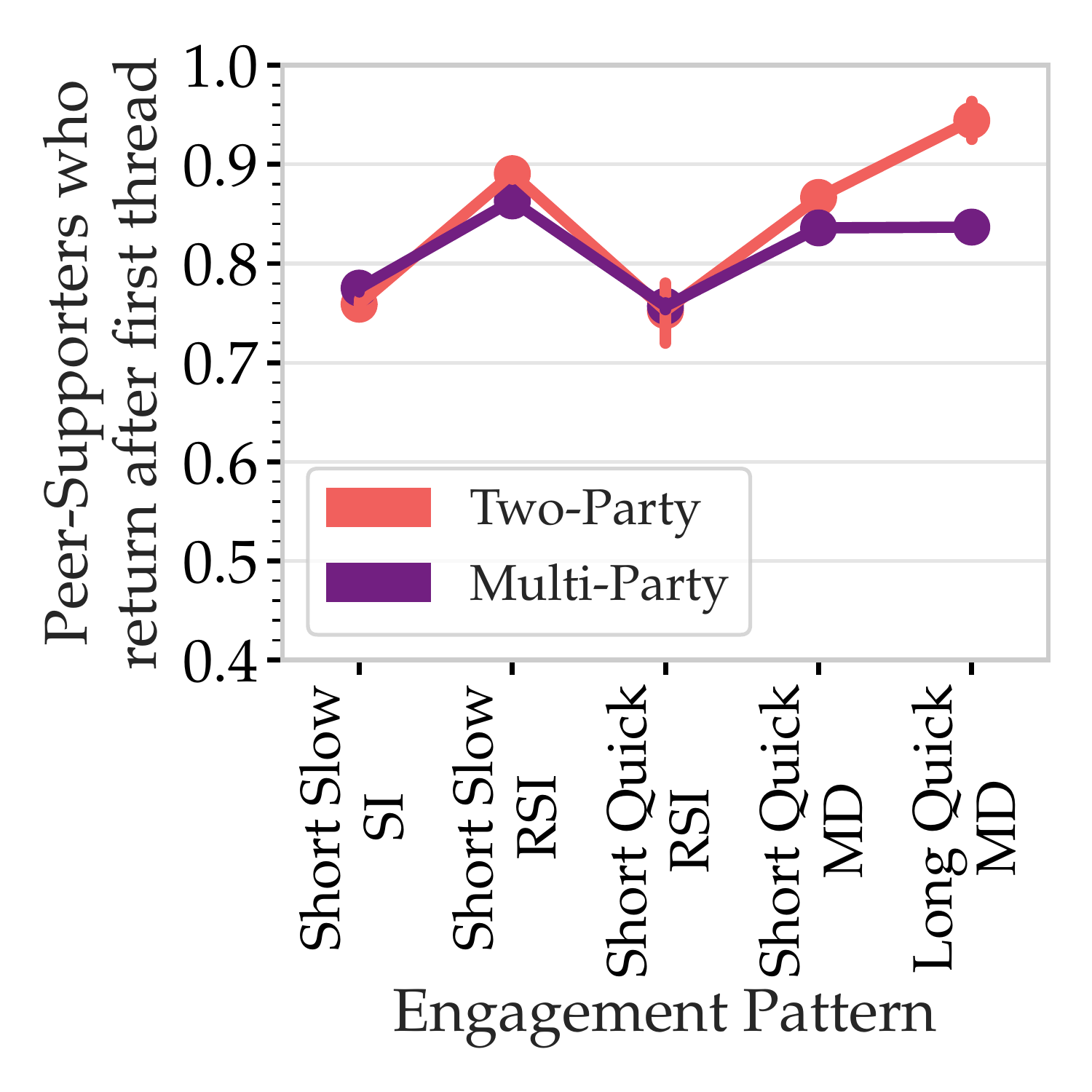} } 
\hfill
\subfloat[\textsc{Reddit}]{
	\label{fig:ps-retention-joint-reddit}
	\includegraphics[width=0.47\columnwidth]{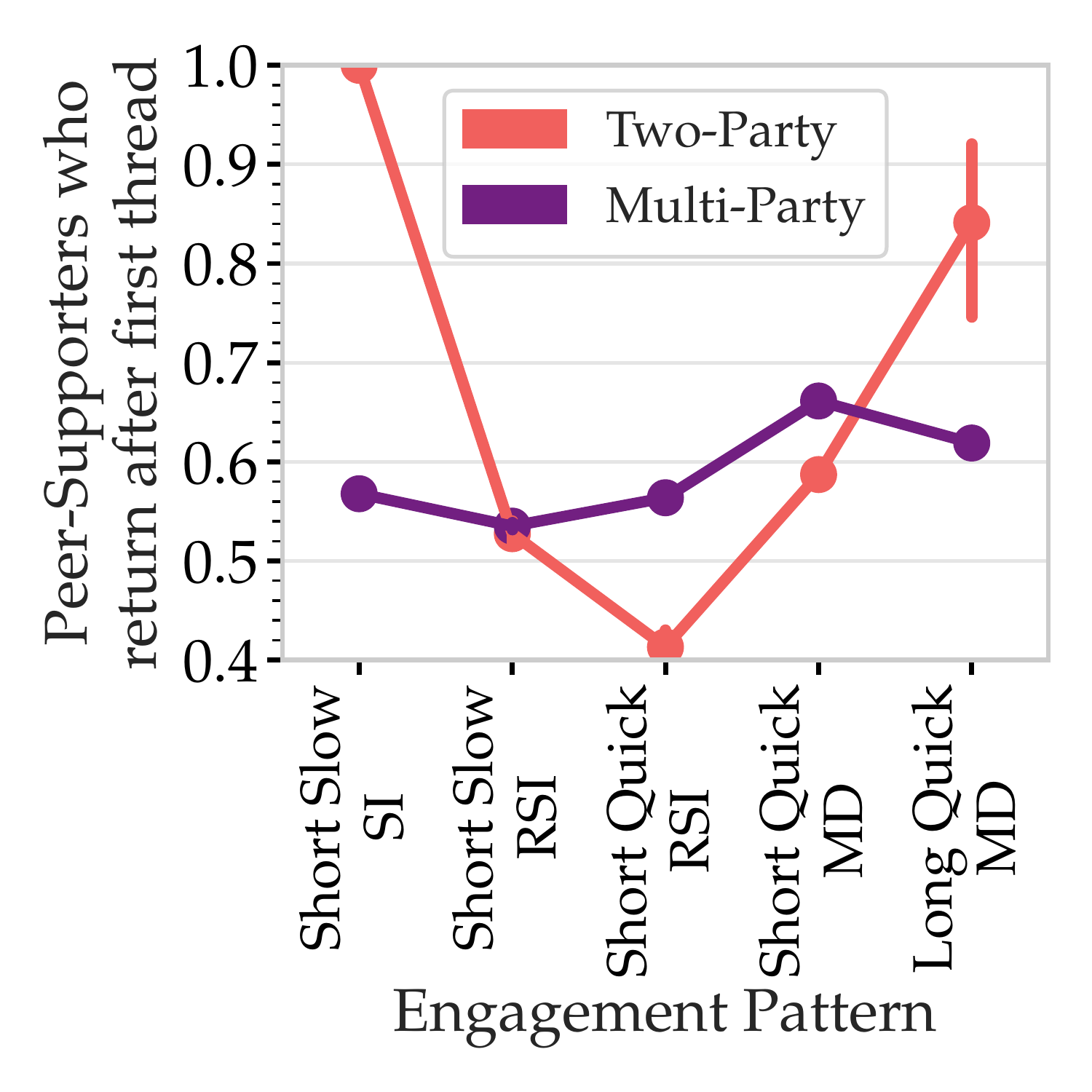} } 

\caption{Peer-supporter retention and engagement patterns. On both platforms, the joint presence of \texttt{Long}, \texttt{Quick} and \texttt{Mutual Discourse} indicators elicits differences between \texttt{Two-Party} and \texttt{Multi-Party} patterns.}
\label{fig:ps-retention-joint}
\vspace{-15pt}
\end{figure}

\xhdr{Peer-Supporters return more often if they were the sole supporters} Analyzing the number of peer-supporters in a thread, we find that \texttt{Two-Party} threads have more peer-supporter retention likelihood than \texttt{Multi-Party} threads on \textsc{TalkLife} (0.81 vs. 0.79; p $<$ 0.01\footnote{Throughout the paper, we use Welch's t-test for statistical testing unless stated otherwise.}). The differences between these two threads are much more prominent in the case of \texttt{Long Quick Mutual Discourse} which is also visible for \textsc{Reddit} (Figure~\ref{fig:ps-retention-joint}; \textsc{TalkLife} - 0.95 vs. 0.84; \textsc{Reddit} - 0.84 vs. 0.62; p $<$ 0.001); the greater differences highlight the importance of looking jointly at the indicators. These observations indicate that a peer-supporter is more likely to {return if they previously were the only supporter in a thread}. This matches the findings in previous work in the context of online crowdfunding, where donors were more likely to return if they were the only donor or one of the very few, presumably due to a stronger sense of personal impact (Althoff et al.~\shortcite{althoff2015donor}). 

\xhdr{Peer-Supporters who are slower-to-act are more likely to return} In our analysis of peer-supporter retention in \texttt{Repeated Seeker Interaction} threads, interestingly, we find that \texttt{Slow} threads have higher likelihood of peer-supporter retention than the \texttt{Quick} threads (Figure~\ref{fig:ps-retention-joint}). These are patterns in which only the seeker interacts repeatedly with the thread. This begs the question of why a \texttt{Slow} thread with seeker response and no second response from the peer-supporter (no \texttt{Mutual Discourse}) will have high correlation with peer-supporter retention. For this, we take a deeper look at the threads of the two types (\texttt{Slow} and \texttt{Quick}) and look at the behavior of seeker and the peer-supporter in the thread. We compare the ratios of peer-supporter's response times and the seeker's response times for the \texttt{Slow} and \texttt{Quick} threads. We find that the peer-supporter's response is, on average, 33 times slower than the seeker's response in \texttt{Short Repeated Seeker Interaction} threads; it is only three times slower in the \texttt{Quick} counterparts. This contrast between the ratios and the corresponding retention likelihoods hint towards associations between response time of peer-supporters and their retention.

In order to better understand the dynamics between the two, we take a closer look at the first-time peer-supporters. We find that retention likelihood of peer-supporters is directly correlated with the response times in their first thread; {a} slow first response has a higher retention likelihood than a quick first response (Figure~\ref{subfig:ps-frt}). This indicates that first-time peer-supporters who are slower-to-act are more likely to return to the platform. These peer-supporters may be acting slowly due to multiple reasons. They might be carefully finding and selecting the threads to respond to; they might be taking more time to write; or they might be getting to know the platform interface. Disentangling these explanations is an important direction for future work.

\subsection{Engaging in \texttt{Mutual Discourse}}
\label{subsec:mutual-discourse}

\begin{figure}
\centering
\subfloat[The likelihood of peer-supporter retention increases if they are slower-to-act. Response time is divided into quartiles in the plot.]{
	\label{subfig:ps-frt}
	\includegraphics[width=0.47\columnwidth]{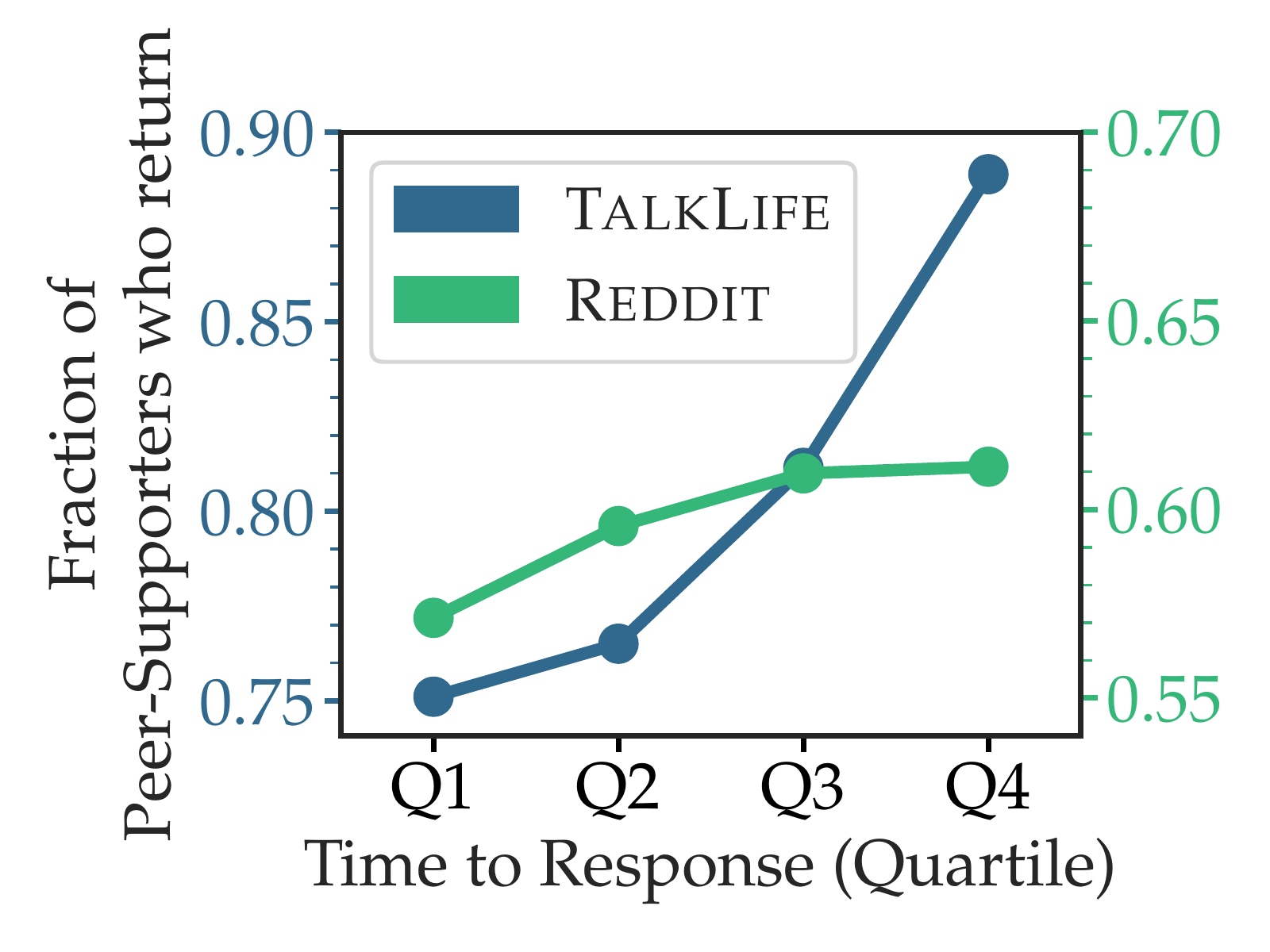} } 
\hfill
\subfloat[Mutual Discourse more likely if the seeker responds early, right after the first peer-supporter.]{
	\label{fig:seeker-position-class-wise}
	\includegraphics[width=0.47\columnwidth]{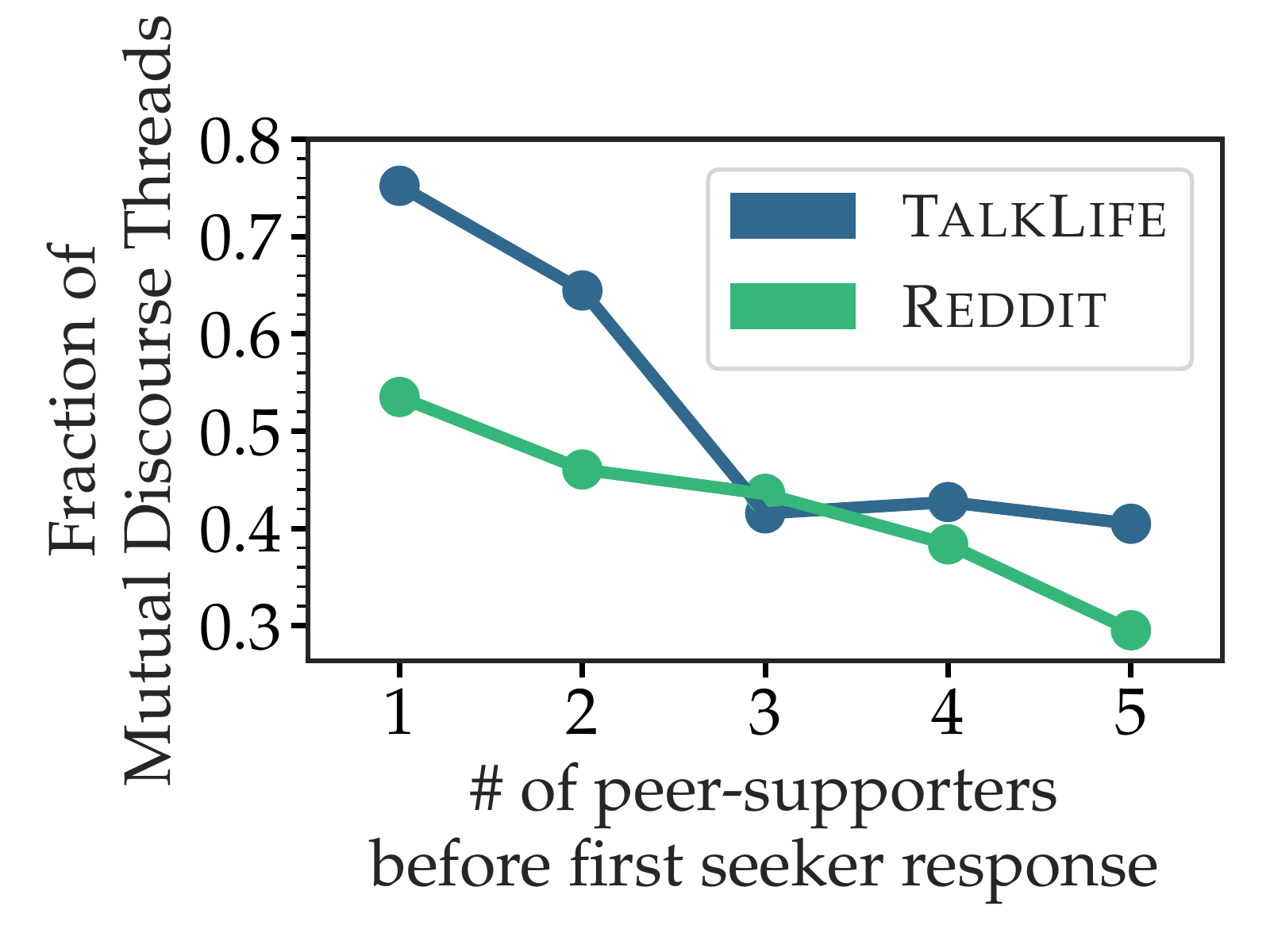} } 

\caption{}
\label{fig:ps-retention-detailed}
\vspace{-15pt}
\end{figure}

We demonstrated that {\texttt{Mutual Discourse} is an important pattern of engagement} associated with higher seeker retention (Section~\ref{subsec:seeker-retention}) and peer-supporter retention (Section~\ref{subsec:ps-retention}) likelihood. 
Next, we investigate when do threads become a \texttt{Mutual Discourse} and what are the factors associated with it. We aim to gain insights on what type of seekers engage repeatedly with peer-supporters and what characteristics of seekers and their posts elicits mutual discourse. Specifically, we focus on threads which evolve into \texttt{Mutual Discourse} after a response from the seeker (the case of \texttt{Repeated Seeker Interaction}). We present our analysis and results next.

\xhdr{\texttt{Mutual Discourse} more likely with seekers who write more and who express negative sentiments in their responses} We analyze the first response of a seeker in a \texttt{Repeated Seeker Interaction} thread and compare it with the responses in \texttt{Mutual Discourse}. We extract the top phrases in both these type of responses separately using TopMine~\cite{el2014scalable}. We observe that responses by seekers in \texttt{Mutual Discourse} contain phrases having more negative sentiment associated with them (e.g. \textit{feel like shit}, \textit{commit suicide}, \textit{don't have friends}) relative to \texttt{Repeated Seeker Interaction}. We quantitatively analyze sentiment of the seeker responses using VADER~\cite{hutto2014vader}. We find that the average sentiment in \texttt{Mutual Discourse} is more negative (\textsc{TalkLife} - 0.096 vs. 0.074; \textsc{Reddit} - 0.078 vs. 0.064; p $<$ 0.01) and less positive (\textsc{TalkLife} - 0.152 vs. 0.222; \textsc{Reddit} - 0.133 vs. 0.171; p $<$ 0.001) than in \texttt{Repeated Seeker Interaction}. This indicates that seekers who post negative sentiment in their responses are more likely to be involved in \texttt{Mutual Discourse}. We also find that the seeker responses in \texttt{Mutual Discourse}, on an average, contain more words (17.16 vs. 13.89; p $<$ 0.001).

\xhdr{\texttt{Mutual Discourse} more likely when seekers respond early} {Next, we investigate when do seekers respond in a thread and if it correlates} with the thread evolving into a \texttt{Mutual Discourse}. We find that threads where seekers respond right after the first peer-supporter are more likely to be \texttt{Mutual Discourse} (Figure~\ref{fig:seeker-position-class-wise}; 75.22\% \& 53.51\% for \textsc{TalkLife} \& \textsc{Reddit} respectively). The likelihood tends to decrease if more number of peer-supporters reply in between. This can potentially be useful in designing support interventions in which the seeker is persuaded to respond early so that a \texttt{Mutual Discourse} is possible.


\section{Conclusion}
\label{sec:discussion}

\xhdr{Summary} Online peer-to-peer support platforms facilitate mental health support but require users on the platforms to interact and engage. In this paper, we conducted a large-scale study of thread-level engagement patterns on two mental health support platforms, \textsc{TalkLife} \& \textsc{Reddit}. We operationalized four theory-motivated engagement indicators which were then synthesized into 11 distinct, interpretable patterns of engagement using a generative modeling approach. Our framework of engagement is multi-dimensional and models engagement jointly on the amount of the attention received by a thread and the interaction received by the thread. We then demonstrated how our framework of engagement patterns can be useful in evaluating the functioning of mental health platforms and for informing design decisions. We contrasted between \textsc{TalkLife} and \textsc{Reddit} using their engagement pattern distributions which suggested that topically focused sub-communities, as found on \textsc{Reddit}, may be important in making online support platforms more engaging. We found \texttt{Mutual Discourse} to be critical for seeker retention, particularly on \textsc{TalkLife}, facilitating which forms an important immediate future research direction.

\xhdr{Risks and limitations} Our study provides new insights on the patterns of engagement between seekers and peer-supporters and the associations of these patterns with their retention behavior. While these insights may have direct implications on the design of mental health support interventions, we note that our analysis is correlational and we cannot make any causal claims. Future work is needed to investigate the causal impact of engagement patterns and their effects on short-term and long-term individual health~\cite{saha2020causal}. Also, our framework of engagement does not account for content of posts, using which often involves ethical risks. This restricts the usage of certain popular interaction theories (e.g.~\cite{sheizaf1988interactivity}) in which the process of authoring of a post is dependent on the content of the previous posts in the thread. 

Finally, we recommend that researchers and platform designers carefully consider the associated risks when considering interventions. For example, we found that \texttt{Mutual Discourse}, which has strong associations with both seeker and peer-supporter retention, is usually characterized by seekers replying back early to the thread. This may prompt platform designers to build, say an app which persuades support seekers to post early responses. However, such an intervention may have unintended consequences since we also found that the responses in \texttt{Mutual Discourse} often involve negative sentiment. Thus, persuading  seekers to repeatedly report self-disclosures~\cite{de2014mental,yang2019channel} with negative sentiments may risk inducing negative effects on a potentially vulnerable population.

\section*{Acknowledgments}
We would like to thank TalkLife for providing us access to the licensed data and Sachin Pendse for help with initial data processing. We also thank Taisa Kushner, Mike Merrill, and Koustuv Saha  for their feedback on this work. Tim Althoff was funded in part by NSF grant IIS-1901386 and the Allen Institute Institute for Artificial Intelligence.

\bibliography{ref}
\bibliographystyle{aaai}

\end{document}